\documentclass[preprint]{aastex}

\usepackage[numbers]{natbib}
\slugcomment{Edit 7 May 2010 v3}
\shorttitle{Magnetic Field in Orion BN/KL} \shortauthors{Tang et al.}

\begin{document}
 
\title{High-Angular Resolution Dust Polarization Measurements: Shaped B-field 
Lines in the Massive Star Forming Region Orion BN/KL}

\author{Ya-Wen Tang}
\affil{Institute of Astronomy and Astrophysics, Academia Sinica,
Taipei, Taiwan}
\email{ywtang@asiaa.sinica.edu.tw}

\author{Paul T. P. Ho}
\affil{Institute of Astronomy and Astrophysics, Academia Sinica, Taipei,
Taiwan \& \\ Harvard-Smithsonian Center for Astrophysics, 60 Garden Street, Cambridge, MA 02138,
USA}
\email{pho@asiaa.sinica.edu.tw}

\author{Patrick M. Koch}
\affil{Institute of Astronomy and Astrophysics, Academia Sinica,
Taipei, Taiwan}
\email{pmkoch@asiaa.sinica.edu.tw}

\author{Ramprasad Rao}
\affil{Institute of Astronomy and Astrophysics, Academia Sinica, Taipei,
Taiwan}
\email{rrao@asiaa.sinica.edu.tw}

\begin{abstract}
We present observational results of the thermal dust continuum emission and its 
linear polarization in one of the nearest massive star-forming sites Orion BN/KL in 
Orion Molecular Cloud-1. The observations were carried out with the Submillimeter 
Array. With an angular resolution of 1$\arcsec$ ($\sim$2 mpc; 480 AU), we have detected and 
resolved the densest cores near the BN/KL region. At a wavelength of 
$\sim$870 $\mu$m, the polarized dust emission can be used to trace the structure of the magnetic field in this 
star-forming core. The dust continuum appears to arise from a $V$-shaped region, with 
a cavity nearly coincident with the center of the explosive outflows observed on larger 
scales. The position angles (P.A.s) of the observed polarization vary significantly by 
a total of about 90$\degr$ but smoothly, i.e., curl-like, across the dust ridges. Such a polarization pattern can be explained with dust grains being magnetically aligned instead of mechanically with outflows, since the latter mechanism would cause the P.A.s to be parallel to the direction of the outflow, i.e., radial-like. The magnetic field projected in the plane of 
sky is therefore derived by rotating the P.A.s of the polarization by 90$\degr$. We find an azimuthally 
symmetric structure in the overall magnetic field morphology, with the field directions 
pointing toward 2$\farcs$5 west to the center of the explosive outflows. We also find a preferred symmetry plane at 
a P.A. of 36$\degr$, which is perpendicular to the mean magnetic field direction (120$\degr$) of the 
0.5 pc dust ridge. Two possible interpretations of the origin of the observed magnetic field structure are discussed. 
\end{abstract}


\section{Introduction}

Massive star-forming sites are typically further away than low mass cases are. Moreover, 
massive stars are formed in groups and are almost always in a more complex environment (Lada \& Lada 2003). Therefore, observations with higher spatial resolution are vital in order to study the 
massive star formation process. As one of the closest massive star-forming sites, with a size scale $>$ 0.07 pc the Kleinmann$-$Low (KL; Kleinmann \& Low 1967) nebula in the 
Orion molecular cloud is the 
best candidate for study. In this paper, we adopt a distance of 480 pc to the KL nebula (Genzel 
et al. 1981). Within the KL nebula region, there are many young stars. The Becklin$-$Neugebauer 
(BN) object is the brightest near-infrared source (Scoville et al. 1983). The source IRc2 is strong 
in mid-IR emission and has been resolved into several isolated peaks (Shuping, Morris \& Bally 
2004), suggesting that it is a cluster-forming region. Using high angular resolution (up to 0$\farcs$7) sub-millimeter (sub-mm) observations, Beuther et al. (2004; 2006) further identified the sub-mm 
counterpart of the radio source I and a new sub-mm source SMA 1. The reported gas mass of source 
I as estimated from the sub-mm continuum is $\sim$0.2 M$_{\sun}$, while the mass of SMA 1 could not be inferred from their data because of line contamination; almost no line-free channels were found 
(Beuther et al. 2004). 

Some of the stars are young enough that accretion and outflows can be detected. Reid et al. (2007) reported the existence of an ionized accretion disk surrounding source I with a radius of 
35 AU at a position angle (P.A., positive from the north to the east) of $\sim - $40$\degr$. Plambeck et al. (2009) 
detected a bipolar molecular outflow along a NE$-$SW axis as traced by the SiO $v$=0 $J$=2-1 line which 
originates from source I. The orientations of this accretion disk and the outflow are consistent 
with the observations of the proper motion of the SiO masers (Chandler 2005; Goddi et al. 2009). 
In contrast, the high velocity outflow detected in CO (Zapata et al. 2009; Chernin \& Wright 1996) 
and in H$_{2}$ (Kaifu et al. 2000; Nissen et al. 2007) are in the NW$-$SE direction and are extended over much larger scales of 0.1$-$0.2 pc. These outflows at different size scales are most likely driven by different sources 
and ejected at different times. 

The stellar motions suggest a different kind of outflow. Based on the proper motions of BN, 
source I, and source n, which are on the order of 10 mas yr$^{-1}$ , Gomez et al. (2005) have proposed 
that they are run-away stars from a common center, somewhere between BN and source I about 
500 yr ago. Hereafter, we will refer to this common center as the BN-I center. The adopted coordinate of the BN-I center is: R.A.(J2000) = 05$^{\rm h}$35$^{\rm m}$14$^{\rm s}$.35, decl(J2000) =$-$5\degr22\arcmin27$\farcs$7. Recent observations of 
the large-scale CO outflow by Zapata et al. (2009) suggest that it is linked to the proper motions 
of BN, and the sources I and n. Associated with BN/KL is a hot molecular cloud core with 
a size of 0.01 pc (Beuther et al. 2006). Observed with NH$_{3}$ $J$=3-2 with an angular resolution of 
1$\arcsec$, the linewidths of the individual clumps in the Orion BN/KL core are narrow, smaller than the 
velocity resolution of 1.3 km s$^{-1}$ in most clumps (Migenes et al. 1989). However, the relative 
peak velocities of the clumps are much larger than the thermal motions. This supersonic motion 
between clumps could be due to H II expansion, outflow motions, Alfv\'{e}nic motions 
driven by the $B$ field, or other systematic motions.

The polarization of the emission from the Orion Nebula has been observed at various wavelengths and on various size scales. At a 0.5 pc (3$\farcm$5) scale, the polarization at 350 and 450 $\mu$m seems 
to be uniform, and the magnetic ($B$) field directions are perpendicular to the dust ridge at that scale (Houde 
et al. 2004; Vaillancourt et al. 2008). Schleuning (1998) proposed that the $B$ field starts to exhibit an
hourglass-like morphology at a 0.5 pc scale based on the 100 $\mu$m polarization results. Observed 
with a higher angular resolution of 0.01 pc (4$\arcsec$) at 1.3 mm and 3 mm with the BIMA interferometer, 
the P.A.s of the polarization were found to vary significantly on the 0.02 pc (10$\arcsec$) scale (Rao et al. 1998). 
Observations with 1$\arcsec$ resolution at mid-IR by Aitken et al. (1997) also showed a significant change 
in P.A.s near IRc2. Similar structures in the polarization were also reported by Chrysostomou et al. (1994) using the H$_{2}$ $\nu =$ 1-0 S(1) line, which traces hot gas originating primarily in shocks. In contrast to the mid-IR results, the polarized emission observed at sub-mm wavelengths can directly trace the dense 
regions, where the stars are forming. High angular resolution observations of the polarization at 
sub-mm can therefore be crucial to understanding the $B$ field structure near IRc2.

In this paper, we present $\lambda \sim$ 870 $\mu$m observations of the Orion BN/KL region. By combining 
three different array configurations, we are able to achieve a much better $uv$ coverage than the 
observations by Beuther et al. (2004), which also had a comparable angular resolution of 0$\farcs$8. We 
detect dust structures $>$ 0.01 M$_{\sun}$ (see Section 3.1), with size scales ranging 
from 1.7 mpc to 0.19 pc. We compare the mass distribution and dynamics of these structures 
with the $B$ field morphology as deduced from the dust polarization maps.

\section{Observation and Data Reduction}

The observations were carried out on 2008 Jan 6, and 2009 Jan 25 and Feb 15 using the Submillimeter Array (SMA; Ho, Moran \& Lo 2004)\footnote{The SMA is a joint project between the
Smithsonian Astrophysical Observatory and the Academia Sinica
Institute of Astronomy and Astrophysics and is funded by the
Smithsonian Institution and the Academia Sinica.} in the compact, subcompact, and extended array 
configuration, respectively. The projected baseline lengths ranged from 7 to 253 k$\lambda$ ($\lambda$ = 870 $\mu$m), corresponding to structures from 0$\farcs$8 to 80$\arcsec$ in size. The phase center was at R.A.(J2000) = 5$^{\rm h}$35$^{\rm m}$14.5$^{\rm s}$, decl(J2000) = -5\degr22\arcmin30$\farcs$4. Other observational details are listed in Table 1. The SMA receiver system had two sidebands, each with a bandwidth of $\sim$2 GHz, when the observations were carried out. 
The sampled sky frequencies range from 345.5 to 347.5 GHz in the upper sideband and from 335.5 
to 337.5 GHz in the lower sideband, with a uniform spectral resolution of 0.812 MHz (corresponding 
to a velocity resolution of 0.7 km s$^{-1}$ ). At these frequencies, the primary beam (field of view) of 
the SMA is 32$\arcsec$. Within the observational bandwidth, there is a significant contribution to the 
total emission from spectral lines of a number of molecular transitions (notably CO $J$=3-2 and SiO 
$J$=8-7). The continuum is generated after removing the spectral line contamination. The relative 
pointing accuracy of our maps is $\sim$0$\farcs$06, which is estimated from the mean phase noise of 10$\degr$ of the gain calibrators within the synthesized beam of 1$\arcsec$. 

The angular resolution (synthesized beam) $\theta_{\rm syn}$ achieved depends on the weighting of the 
visibilities used to make the map. Using natural weighting of the visibilities combined from all three tracks, the $\theta_{\rm syn}$ is 1$\farcs$2 $\times$ 1$\farcs$1. The noise level of the Stokes $I$ image $\sigma_{\rm I}$ is 90 mJy beam$^{-1}$, which is 14 times the theoretical expectation. The reason for the degraded $\sigma_{\rm I}$ is the presence of bright large-scale structures, which could not be recovered because of the lack of short spacings. The noise levels of both the Stokes $Q$ and $U$ images, which are much fainter, are much closer to the theoretical noise level, at 9 mJy beam$^{-1}$. For the maps obtained from the extended array data with natural weighting, $\sigma_{\rm syn}$ is 0$\farcs$8 $\times$ 0$\farcs$7 and $\sigma_{\rm I}$ is  40 mJy beam$^{-1}$. The presented maps are made by combining different tracks with different visibility weightings in order to show both the high-resolution compact structures as well as large-scale extended structures. We note that these presented results are one of the highest spatial resolutions and the most sensitive synthesis maps of the polarization properties of thermal dust continuum at 870 $\mu$m (\S 3.2) toward Orion BN/KL at the moment. 

The data were calibrated and processed using the software package MIRIAD (Wright \& Sault 
1993). The gain calibration was obtained from observations of the quasar 0528+134, 0423$-$013, or 
0510+180 (see Table 1). It is necessary to remove the contributions due to instrumental polarization 
(leakage), as these are roughly similar in magnitude to the observed source polarization and can 
corrupt the data (see Marrone et al. 2006 and Marrone \& Rao 2008 for the details of this method). 
The instrumental polarization was obtained from observations of the strong quasar 3c273 for 2 
hr during transit in each track. The leakage term for each antenna and each sideband is 
derived independently. The total intensity (Stokes $I$) map was deconvolved using the task CLEAN 
in MIRIAD. When deriving the polarized intensity ($I_{\rm p}$) and P.A.s with the task 
IMPOL, we used the dirty maps of Stokes $Q$ and $U$ instead of the cleaned ones. This was done 
in order to avoid any possible bias that could be introduced by the CLEANing process. The task 
IMPOL also removed the effects of the bias of the positive measure of $I_{\rm p}$. The noise level of $I_{\rm p}$, $\sigma_{\rm p}$ is 9, 12, and 8 mJy beam$^{-1}$ for the maps obtained from the combined three tracks, from the combined compact and subcompact tracks, and from the extended track, respectively. \textit{Except in Section 4.3, the P.A.s of polarization and the $B$ field throughout this paper are defined to start from the north and increase counter-clockwise from 0$\degr$ to 180$\degr$.} The uncertainties of the detected polarization P.A.s are in the range of 2$\degr -$16$\degr$.

\section{Results}
\subsection{Continuum Emission}

The 870 $\mu$m continuum emission in Orion KL is resolved into
three clumps (Figure 1) which form a $V-$shape. Note that nearly all of the 870 $\mu$m emission is due 
to thermal dust emission (later in this section). 
The brightest dust continuum clump is associated with
source I and SMA 1; hereafter, we call it the \textit{main dust
ridge}. The continuum peak is 1$\arcsec$ southeast of source I
and 2$\arcsec$ north-east of SMA 1. The infrared sources IRc2 are
to the northwest of source I, again offset from the peak of the
main dust ridge. With the highest angular resolution
obtained with the data from the extended array with natural
weighting (Figure 1(b)), the continuum peaks near source I and SMA 1 were
resolved. The position of SMA 1 is offset from the one reported by
Beuther et al. (2004) by 0$\farcs$5. This could be due to the difference in the deconvolution process or difference in the $uv$ coverage. Nevertheless, our results confirm the
detection of the peaks of source I and SMA 1 by Beuther et al.
(2004). Besides, we also find that SMA 1 is embedded in the larger structure
associated with the hot core.

In the west, two fainter dust clumps are detected to the south of BN. Note that the sub-mm 
counterpart of BN is clearly detected in our maps generated from the longest baselines only but not from the 
combined tracks. This could be due to the limitation of the dynamical range in the combined tracks, where bright and extended structures are recovered. The southwest clump is situated at the 
so-called Ócompact ridgeÓ, where strong line emission of oxygen-bearing molecules was detected 
(Beuther et al. 2005). The infrared sources IRc4 and IRc5 are in the periphery of the compact 
ridge. The northwest clump is near the infrared sources IRc3, IRc6, and IRc20. Note that all the 
infrared sources are offset from the dust continuum peaks. Such offsets suggest that these sources 
are most likely physically located outside of the dense dust continuum clumps. Measurements of 
the extinctions toward these sources are required to estimate their relative positions along the line 
of sight, with respect to the dust emission peaks. 

The detected flux density in the main dust ridge, compact
ridge, and the northwest clump is 22$\pm$2 Jy, 8$\pm$1 Jy, and
8$\pm$1 Jy, respectively. When measured with the James Clerk Maxwell telescope, the
brightest point with 14$\arcsec$ angular resolution is 132 Jy
beam$^{-1}$ (Vall\'{e}e \& Fiege 2007). The reported SMA measurement
therefore recovers 29\% of the total flux density. When fitted
with Gaussians toward the emission peaks in Figure 1(b), the
detected flux density in BN, Source I, and SMA 1 is 2.2$\pm$0.2 Jy, 1.4$\pm$0.1 Jy, and
3.3$\pm$0.3 Jy, respectively. Note that the free$-$free emission in the region is very faint, being 
4.5, 1.1, and 2 mJy at 3.6 cm and 31, 13, and $<$4 mJy at 7 mm for BN, source I, and source n, respectively (Menten 
\& Reid 1995). The estimated free$-$free emission toward source I is in the range of 44$-$400 mJy at 870 $\mu$m, depending on the emission sources and physical conditions in source I (Beuther et al. 2006; Reid et al. 2007). Compared to the flux density detected toward source I within a $\sim$ 1$\arcsec$ region, 
the free$-$free contribution is in between 3 and 30\%. It is $<$2\% when compared to the main dust ridge, which is $\sim$ 10$\arcsec$ in length.

There is no 870 $\mu$m continuum peak detected at source n
and the BN-I center. Furthermore, the morphology of the 870 $\mu$m
dust continuum emission is similar to the NH$_{3}$ line emission
structure as seen by Migenes et al. (1989) and Wilson et al.
(2000), i.e., being bright to the east of source I, source n, and
SMA 1, and with a cavity to the west of these sources. The
coincidence with the NH$_{3}$ line emission suggests that the revealed
870 $\mu$m continuum is mostly tracing the dense gas. Due to the
significant differences in the spatial distributions of the
molecules detected in these regions, Blake et al. (1987) and
Beuther et al. (2005) suggest that these molecular line maps may
help to differentiate evolutionary stages. The importance of the
870 $\mu$m dust emission is that it is optically thin, impervious
to chemistry and reliably tracks the total mass distribution if dust grains are not destroyed.

Associated with the main dust ridge is a "hot core",
where a lot of hot core lines were detected (Beuther et al. 2005).
The gas temperature is $\sim$ 165$-$400 K (Wilson et al. 2000).
Assuming of local thermal equilibrium, the dust emission traced here is most
likely tracing dense and hot gas. We assume that the dust grains
are at a temperature of 160 K as traced with NH$_{3}$. The gas
mass ($M_{\rm gas}$) in the main dust ridge estimated
from the dust continuum is then 2$-$12 M$_{\sun}$, assuming a gas
to dust ratio of 100, where the limits for $M_{\rm gas}$ are from
the dust emissivity $\beta$ of 1 or 2. Note that the stellar mass
of source I derived based on the kinematics of the maser spots is
6$-$10 M$_{\sun}$ (Greenhill et al. 2004). Based on the
linewidths of 10 NH$_{3}$ clumps, Migenes et al. (1989)
reported the binding mass within each clump using the virial theorem. All of the clumps have a binding mass in the range $<$0.7 M$_{\sun}$ to 17.3 M$_{\sun}$. 
Note that all $M_{\rm gas}$ estimated from the kinematics are larger than our
estimates, e.g., $M_{\rm gas}$ of source I is in the range 0.1$-$0.8 M$_{\sun}$ from the dust continuum but in the range $<$ 0.7 to 17.3 M$_{\sun}$ from the virial estimate.
Possible explanations for this discrepancy might be that the majority of the dust grains we traced have
been dispersed by stellar feedback, the gas
temperature is extremely low in the cores, the gas$-$to$-$dust
ratio is significantly higher than 100, or the NH$_{3}$ clumps are not in virial equilibrium. Based on the same assumptions, $M_{\rm gas}$ in the compact dust ridge and the NW clump are both in the range 0.6$-$4 M$_{\sun}$.
Note that the elongation of the main dust ridge is
perpendicular to the 0.5 pc $B$ field direction in the OMC-1 dust
ridge as traced by Schleuning (1998) at a wavelength of 100
$\mu$m.

\subsection{Polarization}
The linearly polarized emission ($I_{\rm p}$) is detected and resolved in the main dust ridge and in 
the southwest clumps (color scale in Figure 2), and it is barely detected in the northwest clump. 
Near IRc2 to the west of the main dust ridge, there is weak or no $I_{\rm p}$ detected (Figure 2(b)). $I_{\rm p}$ is 
brightest in the south of SMA 1.  

The detected polarization exhibits significant deviations in P.A.s between clumps. Among these, 
the polarization vectors extending from source I toward the south have very different P.A.s from 
the larger scale polarization direction at a mean P.A. of 26$\degr$ detected at a 0.5 pc scale (Vall\'{e}e \& 
Fiege 2007). The rest of the vectors in the main dust ridge are consistent with the larger scale 
polarization direction, indicating that the $B$ field is retained mainly on a larger scale but not in the 
cores. With the combined data from the compact and subcompact arrays, the polarization from 
the extended structures is better detected (Figure 2(c)). The polarization in the northwest clump and 
in between the three clumps is detected. Extending from the northwest clump toward infrared 
source n and then toward the hot core, the polarization direction appears to vary continuously 
and smoothly. The polarization direction in the northwest clump, in the west, and east of the main dust ridge is nearly parallel to the field direction at the 0.5 pc scale. The most striking feature revealed with the 870 
$\mu$m polarization data is that the P.A.s vary smoothly and substantially across the main dust ridge, 
from a P.A. of 10$\degr$ in the northeast to a P.A. of 100$\degr$ in the south. With higher weighting toward 
more extended structures (Figure 2(c)), the variation of polarization appears to be continuous to 
the northwest clump and to the compact ridge. 

\subsection{Comparison of Polarization at 3 mm, 1.3 mm and 0.87 mm}
With BIMA, Rao et al. (1998) measured the polarized emission at 3
mm and 1.3 mm with an angular resolution of $\sim$ 4$\arcsec$ (Figure
3). The polarization at 3 mm and 1.3 mm extends along the major axis
of the OMC-1 dust ridge, where the P.A.s are consistent with the
0.5 pc scale polarization direction. Apart from the components
along the dust ridge, there is one peculiar extension detected at 3 mm from the
main dust ridge toward the southeast, where the
direction of polarization (with P.A. of $\sim$ 70$\degr - \sim$ 100$\degr$) is very different from the ones at
other positions (with P.A. of $\sim$ 30$\degr$, roughly parallel to the 0.5 pc polarization). Note that in Rao et al. (1998), the polarized emission at 3 mm seems
to have a more extended structure as compared to the 1.3 mm map. This
is most likely due to the differences in sensitivities at these
two wavelengths.

With the higher angular resolution of the SMA, the polarizations north of source I and 
near IRc5 are clearly detected. Previous BIMA maps appear to be depolarized here. At these two 
locations, the P.A.s appear to vary significantly within the dust ridge. We have tested the effects 
of differences in resolution by convolving our CLEANed images of Stokes $I$, $Q$, and $U$ maps with 
$\theta_{\rm syn}$ of the BIMA measurements at 3 mm. Most polarization vectors disappeared near IRc5 and north 
of source I. The depolarization previously seen is clearly due to the smearing of a more complex 
underlying structure, alike what we have found in the W51 e2/e8 region (Tang et al. 2009). Since 
the 870 $\mu$m polarization is observed with only a single field, we are able to image the polarization 
only within the field of view of 32$\arcsec$. The reason why there is no polarization detected at 870 $\mu$m 
along the major axis of the OMC-1 dust ridge is mainly due to the decrease of the primary beam 
response near the edge of our field of view. Although the polarizations vary significantly in the 
KL region, our results show that such a variation is smooth and continuous. The extension toward 
the southeast, with a very different polarization direction at 3 mm, can possibly be traced back 
toward the south of the main dust ridge and the eastern compact ridge. In summary, our 0.87 mm 
polarization map reveals consistent polarized emission as compared to that seen at 3 mm and 1.3 
mm. The abrupt change previously seen in P.A.s near IRc5 is shown to be due to the more complex 
underlying structure. 

\section{Discussion}

What is new and important in this experiment?
\begin{enumerate}
    \item The dust continuum emission and its polarized components
    are resolved in a massive star-forming region at the 1$\arcsec$ (
    $\approx$2 mpc = 480 AU) scale for the first
    time.

    \item The detected polarization is in general consistent with and complementary to the 3 mm 
and 1.3 mm measurements by BIMA at 4$\arcsec$ resolution. With higher angular resolution, the 
polarization in the north of IRc2 and adjacent to IRc5 have been detected and resolved in 
our 870 $\mu$m map. In the fainter region, the polarization direction is consistent with the 0.5 pc scale 
polarization direction. However, in the south of SMA 1 extending toward IRc5, the P.A.s are 
nearly perpendicular to the larger scale direction. The striking feature is that this deviation 
is smooth across the dust ridge. 

    \item In the main dust ridge, $I_{\rm p}$ is much lower
    or absent on the western side of the main dust ridge,
    immediately to the west of source I.

\end{enumerate}

Based on these results, in the following sections we discuss possible interpretations of the
polarization properties, together with the known kinematics.

\subsection{Kinematics and $B$ Field}
\subsubsection{Large$-$scale Kinematics and 870 $\mu$m Dust Continuum}
Zapata et al. (2009) noted the comparable energies released in the CO outflows and the ejection 
of the three sources (BN, source I, and source n) from the BN-I center, which are both on the order of 10$^{47}$ 
erg s$^{-1}$. Both the outflows and the ejection phenomenon are estimated to have happened 
$\sim$ 500 yr ago. Moreover, the very special morphologies of this highly impulsive and poorly collimated 
molecular outflow (Figure 3(b)) indicate that this case is very different from the typical bipolar 
outflows associated with the isolated star formation process. They propose that the 0.14 pc CO 
outflow may be the result of the disintegration of a central cloud core or circumstellar disks near 
the BN-I center, when these three sources underwent a close dynamical encounter. 

The fact that the 0.14 pc CO outflow filaments are numerous and
bright preferentially in the northern part of the BN-I center can
be explained by our detected 870 $\mu$m continuum emission.
First, there is no dust continuum emission at the BN-I center.
As discussed in Section 3.1, the similarity between the dust continuum
emission and the integrated NH$_{3}$ line emission suggests that
the central cavity is real, and the gas density here is lower. The
close encounter 500 yr ago might have cleared out the majority
of the molecular gas near the BN-I center. Second, the densest
structures appear to be stronger in the SE of the BN-I center,
where the dust ridges are dense enough to withstand the ionizing
radiation and the shock fronts released in the event of the
close encounter. Note that the gas mass of the large-scale outflow is $\sim$ 10 $M_{\sun}$ (Kwan \& Scoville 1976).
The gas mass within the 14$\arcsec$ beam toward IRc2 is 11$-$100 M$_{\sun}$ as estimated from the single dish 870 $\mu$m continuum measurement by Vall\'{e}e and Fiege (2007) with the same assumptions as in Section 3.1. It is difficult to conclude if there was sufficient gas to truncate the outflows before the majority of the gas was dispersed. Note that the luminous Ori C is 1$\arcmin$ away from source I. Besides, the large-scale jets traced by H$_{2}$ are also less numerous in the southeast. It is therefore less likely that the CO bullets were photo-dissociated by Ori C.

\subsubsection{Clump Outflow Interaction and Estimate of B Field Strength}
The kinematics within the dense dust ridge also provide some clues of the physical conditions in this region. 
Note that the NH$_{3}$ 
clumps are narrow in linewidths, being smaller than the spectral resolution of 1.3 km s$^{-1}$, but 
having a maximum difference in velocity between clumps of $\sim$ 15 km s$^{-1}$
(Migenes et al. 1989; marked in Figure 1(b)). As suggested by the same authors, there must be systematic motions 
driving the motions between the clumps. Based on the comparison of NH$_{3}$ (3,2) maps 
integrated at different velocity ranges, Migenes et al. (1989) suggested that the NH$_{3}$ emission 
traces an incomplete expanding cylinder. 

Besides the above scenario, it is highly possible that the molecular gas and the clumps have been impacted by the outflows which may have driven large relative velocities between clumps. The question then arises as to why the linewidths of individual clumps remain so small. 
Can the existence of a $B$ field help to confine the clumps?
To check if the clumps can be held together by their own gravity in the presence of the outflows, we first
compare the ram pressure, $P_{\rm ram}$, and the gravitational binding force per unit area, 
$P_{\rm grav}$, of the clumps. $P_{\rm ram}$ is given by $\rho_{\rm outflow}v{_{\rm rel}}^{2}$, 
where $\rho_{\rm outflow}$ is the mass density of the outflow, and $v_{\rm rel}$ is the relative velocity between the 
NH$_{3}$ clumps and the outflows. No exact value for $\rho_{\rm outflow}$ is found from the literature for this region. We assume that the mass of 10 M$_{\sun}$ contained in the outflow is uniformly 
distributed within a sphere with a radius of 7$\arcsec$.  $\rho_{\rm outflow}$ is then 3.7 $\times$ 10$^{-17}$ 
g cm$^{-3}$, which translates into a number density $n_{\rm H_2} \approx$  2 $\times$ 10$^{7}$ cm$^{-3}$. The outflows detected closest to the BN-I center in Zapata et al. (2009) start from the radial velocity with respect to the local standard of rest, $v_{\rm lsr}$, of 40 
and $-$30 km s$^{-1}$ for the redshifted and blueshifted parts, respectively. Since the NH$_{3}$ clumps are at $v_{\rm lsr}$ of $-$5 to 10.6 km s$^{-1}$ (Migenes et al. 1989), $v_{\rm rel}$ is then in the range between 
25 and 45 km s
$^{-1}$. The estimated $P_{\rm ram}$ is $\sim$ 2 - 7 $\times$ 10$^{-4}$ dyn cm$^{-2}$. $P_{\rm grav}$ is equal to 2$\pi$G$\Sigma_{\rm gas}$$\Sigma_{\rm G,total}$, where $
\Sigma_{\rm gas}$ and $\Sigma_{\rm G,total}$ are the surface mass density of the gas and of the NH$_{3}$ clumps, respectively. From Migenes et al. (1989), $n_{\rm H_2}$ of the NH$_{3}$ clumps is $\sim$10$^{8}$ 
cm$^{-3}$, which gives the mass density of the clump $\rho_{\rm clump} \approx$ 4.3 $\times$ 10$^{-16}$ g cm
$^{-3}$.  From the same reference, the diameters of the clumps are all $\sim$ 1$\arcsec$, which gives a mean $
\Sigma_{\rm G, total}$ of 3 g cm$^{-2}$. Assuming $\Sigma_{\rm gas}$ = $\Sigma_{\rm G, total}$, 
$P_{\rm grav}$ is 3.9 $\times$ 10$^{-6}$ dyn cm$^{-2}$.
Thus, $P_{\rm grav}$ is 1$-$2 orders of magnitude smaller 
than $P_{\rm ram}$, indicating that these clumps would not be able to withstand the ram pressure. Consequently, over time they would disintegrate. This would likely result in broader linewidths than what is observed.

The major uncertainty in the estimated $P_{\rm ram}$ is probably $\rho_{\rm outflow}$. 
If $\rho_{\rm outflow}$ is 2 orders 
of magnitude smaller, being 3.7 $\times$ 10$^{-19}$ g cm$^{-3}$, the clumps can still remain confined. Otherwise another force is needed to sustain the 
clumps. One possible force is the $B$ field surface tension between the outflows 
and the clumps. This tension, in addition to self-gravity, can maintain the clump shape and help to suppress surface 
instabilities (Kevin$-$Helmholtz instability) resulting from ram pressure with sufficient velocity shear. 
Following  Vikhlinin, Markevitch, \& Murray (2001), if $B_{\rm outflow}^{2}$ + $B_{\rm clump}^{2} >$ 4$\pi 
\frac{\rho_{\rm outflow} \rho_{\rm clump}}{\rho_{\rm outflow} + \rho_{\rm clump}}v_{\rm rel}^{2}$, the clumps can 
remain 
confined. Assuming $\rho_{\rm outflow}$ is 3.7 $\times$ 10$^{-17}$ g cm$^{-3}$, the estimated latter term is 4 $
\times$ 10$^{-4}$ erg cm$^{-3}$. Therefore, if $B_{\rm outflow}$ = $B_{\rm clump} \ge$ 8 mG, 
the clumps can remain confined. If $\rho_{\rm outflow}$ is 10 times smaller, being 
3.7 $\times 10^{-18}$ g cm$^{-3}$ ($n_{\rm H_2} \sim 2 \times 10^6$ cm$^{-3}$), 
the $B$ field strength needs to be $\ge$ 3 mG. On the other hand, if $\rho_{\rm outflow}$ is 10 times larger - because 3.7 $\times$ 10$^{-17}$ g cm$^{-3}$ is also possibly a lower limit because the outflows are not uniformly distributed over a sphere -  a $B$ field strength of 30 mG is needed. 

In summary, in order to confine the NH$_{3}$ clumps in the presence of the outflows with $v_{\rm rel}$ of 25$-$45 
km s$^{-1}$ and $n_{\rm H_{2}}$ of the outflows $\ge$ 2$\times$10$^{6}$ cm$^{-3}$, the $B$ field strength 
within the NH$_{3}$ clumps or the outflows needs to be larger than 3 mG. The estimated lower limit of 3 mG is 
roughly consistent with the observed $B$ field strength from OH masers (see Section 4.3.2) of 1$-$16 mG. 
Larger $B$ field strengths, which are required because of larger $\rho_{\rm outflow}$, might be 
achievable through local field compression.
The method presented here provides an independent estimate of the total $B$ field strength in this region.

\subsection{Dust Grains Being Mechanically or Magnetically Aligned?}
There are several molecular outflows detected in the Orion BN/KL region. The large$-$scale CO 
outflows (0.08 pc) are in the northwest$-$southeast direction (Chernin \& Wright 1996). As observed 
with a 3$\arcsec$ resolution and a larger field (0.14 pc), the outflow is more spherically symmetric centered 
at the BN-I center (Zapata et al. 2009; Figure 3(b)). The velocities of the outflows are up to $\pm$100 
km s$^{-1}$. The low velocity outflow centered on source I is in the NE$-$SW direction. An ionized 
flow has been proposed centered on source n in nearly a north$-$south direction (Greenhill et al. 2004; 
Menten \& Reid 1995). If the dust grains are mechanically aligned by these outflows, the detected 
polarization should be nearly parallel to the direction of these outflows due to the Gold alignment 
(Gold 1952), where the elongated dust grains have their major axis aligned with the flow direction 
due to dynamical interactions with the outflowing gas. In this case $-$ with a dominating mechanical 
alignment $-$ the detected coherent polarization is not related to any $B$ field. Although the driving sources 
of these outflows are different, they are located within 10$\arcsec$ from the BN-I center. In the case of 
dust grains that are mechanically aligned, the directions of the polarization P.A.s should be nearly 
radial around the center of the spherical outflows detected by Zapata et al. (2009). No $B$ 
field is needed in this scenario. As shown in Figure 3 and in the right panel of Figure 4, the polarization segments appear to be curl-like, rather perpendicular to the explosive outflow. Additionally, the smooth variation in P.A.s as a function of azimuth (see discussions in Section 4.3) across the dust ridge seems to be difficult 
to explain by the alignment via outflows. Therefore, a mechanical alignment by outflows seems unlikely. 

Based on the theoretical calculations on magnetic alignment mechanisms, it has been suggested 
that dust grains are aligned by the $B$ field in most cases (Lazarian 2007). We will adopt this in the 
following section (Section 4.3) when discussing the polarization patterns as a result of magnetic alignment. 
The direction of the projected $B$ field can then be derived by rotating the detected polarization 
angles by 90$\degr$.

\subsection{Interpretation of the Polarization Resulting from Magnetic Alignment $-$
Possible Overall $B$ Field Morphologies}

Before discussing two possible interpretations of our higher resolution $B$ field morphology, we first summarize the large-scale field geometry. 
At the 0.5 pc scale observed with a 15$\arcsec$ resolution, the $B$ field appears to be uniform across OMC-1. 
Schleuning (1998) further proposed that the $B$ field is Óhourglass-likeÓ along the OMC-1 ridge. The 
same author further pointed out that there is no local pinched field detected near the BN/KL region. 
When zooming in with a 4$\arcsec$ angular resolution, Rao et al. (1998) showed that the polarization nearly 
20$\arcsec$ away from IRc2 is consistent with the 0.5 pc field direction. In the same paper, the polarization 
near BN/KL was found to change abruptly. With a 1$\arcsec$ resolution, we find that the field morphology 
near IRc2 varies significantly but smoothly. Besides, the $B$ field direction away from IRc2 appears 
to be consistent with the 0.5 pc large-scale field direction. In summary, the $B$ fields traced at 0.5
pc (Schleuning 1998), at 0.05 pc (Rao et al. 1998) and at 0.01 pc (this paper) exhibit consistent 
field directions away from IRc2. With a 1$\arcsec$ resolution, the field geometry near the densest region 
BN/KL is clearly resolved. 

Following the argument in Section 4.2, the dust grains are most likely aligned via the $B$ field 
and a mechanical alignment is unlikely. In Figure 4, the $B$ field map is presented by rotating the 
polarization P.A. by 90$\degr$ with identical length for each segment. We note that the field lines detected 
near IRc2 in BN/KL all point toward the BN-I center. In order to further analyze this possible 
correlation, we plot the distribution of the $B$ field P.A.s as a function of the azimuth angle of the emitting region. For this 
purpose, azimuth is defined to increase counterclockwise from 0$\degr$ to 360$\degr$ 
starting from the north. The origin of the azimuth
reference frame is chosen to be 2$\farcs$5 west of the BN-I center, as determined by the best correlation between 
the $B$ field directions and the azimuth angles.  \textit{Note that in order to check the azimuthal symmetry property, the P.A.s of the observed $B$ fields (conventionally defined in a range of 180$\degr$) are re-defined in the same way as the azimuth angle over a range of 360$\degr$. For example, a P.A. of 20$\degr$ for the $B$ field is re-defined as 200$\degr$ if the corresponding azimuth angle is 270$\degr$.} As shown in Figure 5, there is 
a general trend that the observed $B$ field direction varies linearly as a function of azimuth with a 
slope of one. This characterizes the azimuthally symmetric structure in the overall $B$ field morphology, 
especially for the data points close to the reference center and with stronger intensity. If the 
origin of the azimuth reference frame is set to the BN-I center, there is a systematic shift of data 
points and therefore, the correlation with azimuth angle is less pronounced.

Nevertheless, given some deviations from the straight line in Figure 5 $-$ especially in the southern region $-$ we further 
check if there is also a preferred (single) azimuth direction possibly revealing a symmetry 
plane. Indeed, we find that there is a preferred symmetry plane (Figure 6), where the corresponding 
$B$ field lines above and below the plane show symmetric structures. As a measure of symmetry, 
we check the rms of the difference in the $B$ field P.A.s, $\triangle$P.A., for corresponding data 
points in the upper and lower symmetry planes. For this purpose, the $B$ field P.A.s above and below 
a possible symmetry plane are re-defined conventionally in the range 0$\degr -$ 180$\degr$ clockwise 
in the upper plane and 
counterclockwise in the lower plane. With this definition, a mirror symmetrical structure will give 
$\triangle$P.A. = 0, and the smaller $\triangle$P.A. the closer the structure is to a mirror symmetry. Since there 
is no exact one-to-one correspondence of the detected polarization in the upper and lower planes, 
we interpolate the data onto a regular grid and then calculate $\triangle$P.A.. The formulation is 
$\triangle$P.A. = $\sqrt{\frac{1}{N}\Sigma^{N}_{i=1} (P.A.^{\rm upper}_{i} - P.A.^{\rm lower}_{i})^{2}}$. For the best-guess symmetry plane along a P.A. of 36$\degr$, $\triangle$P.A. is 
26$\degr$. For a typical non-symmetry plane (right$-$panel in Figure 6), 
$\triangle$P.A. is 48$\degr$. Consequently, the 
symmetry along the plane at a P.A. of 36$\degr$, where $\triangle$P.A. is 26$\degr$, is very significant, minimizing 
$\triangle$P.A. to a level where it is 
close to the measurement uncertainties ($\pm$10$\degr$). We note that this preferred symmetry plane (plane along a P.A. of 36$\degr$) is nearly perpendicular to the large$-$scale $B$ field direction (mean field direction is 
120$\degr$), as might be expected for dense structures forming in the contraction process under a strong 
$B$ field.

Besides the symmetry morphologies discussed above, the P.A. patterns also exhibit remarkably 
smoothly changing features. Such smooth but significant variations in the P.A.s across the dust 
ridges are consistent with the polarization maps obtained in the H$_{2}$ S(1) emission reported by 
Chrysostomou et al. (1994). Again, we emphasize that the sub$-$mm continuum and polarization 
maps are sensitive to the dense structures, while the H$_{2}$ emission traces warm/hot gas with typically $T >$1000 K resulting from shocks (e.g., Kristensen et al. 2007). The consistency between the sub$-$mm and NIR polarizations indicates that the smooth variation in the field directions is continuous, regardless of the 
physical environments. How can the field lines vary so significantly within 0.1 pc? Since the two 
dominant forces in the system are gravity and the explosive event, in the following subsections, we 
discuss two possible shaping mechanisms of the $B$ field morphology where either one of the two 
forces is dominant.

\subsubsection{Poloidal and Toroidal B Fields?}

The $B$ field morphology can originate from collapse and accretion. The partially ionized molecular gas flows along 
the large$-$scale $B$ field, and then dense structures are elongated perpendicular to the field lines. When the dense elongated structures (pseudo-disks) form inside the molecular cloud, a toroidal $B$ field is expected in a magnetized and ionized cloud. The existence 
of a toroidal field component can be due to the differential rotation of the disk (Newman, Newman, 
\& Lovelace 1992), where the $B$ field lines are anchored to the rotating disk such that the $B$ field 
direction is nearly perpendicular to the rotation axis in or near the disk. In this scenario, the $B$ fields in the extension following the OMC-1 dust ridge are tracing the poloidal $B$ field. This is schematically illustrated in 
Figure 7(a). However, there is no evidence of such a 
large$-$scale disk based on our continuum map (Figure 1). As discussed in Section 3.1, the dense structures 
form a V-shape pattern, where the cavity is near the BN-I center. It is possible that this shape 
of dense structures can be formed as a process of ring fragmentation as discussed in Machida et 
al. (2005), although in their model this occurs at a much smaller scale (100 AU scale) and in an 
environment with much higher number density. Alternatively, it is also possible that the disk-like 
structure has been swept away in the event of the close encounter $\sim$ 500 yr ago (Section 4.3.2), still 
possibly preserving poloidal and toroidal $B$ field structures. 

In order to test this scenario, we have to estimate the period of rotation. The rms velocity of the detected NH$_{3}$ emission is 5 km s$^{-1}$ (Migenes et al. 1989). If the radius 
of the pseudo-disk is 10$\arcsec$, as estimated from the scale where the change in polarization direction is 
detected, the period of the rotation is 3 $\times$10$^{4}$ yr. As compared to the time scale of the formation 
process of a typical low$-$mass young stellar object, on the order of Myr, the period is short, which 
means that any disk should have rotated many times. The toroidal field lines embedded in the 
disk would have been dragged for many revolutions and would be wound up. One possibility is 
that ambipolar diffusion allowed the field lines to slip relative to the rotating disk. However, it 
appears contrived to maintain the observed field geometry after many rotations of the disk. It is 
also possible that the field lines are mainly pulled by the contraction process and the accretion 
rate is high enough to overcome the sustaining $B$ field tension. In this scenario, the toroidal field 
structure seen near IRc2 is organized by the contraction process and therefore possibly exhibits an incomplete hourglass structure.  

We also note that an accreting disk for source I has been proposed
(Reid et al. 2007; Matthews et al. 2010), which is the most convincing disk near an
intermediate mass protostar. The disk direction is nearly parallel
to the large$-$scale $B$ field direction, and it is not correlated with
the pseudo$-$disk, indicating that $B$ field directions
are not necessarily retained in the cores due to an event of a recent close encounter (Zapata et al. 2009) or simply due to the difference of angular momentum at various scales (e.g., L1489; Brinch et al. 2007).

\subsubsection{B field lines being shaped and dragged along with explosive
outflows?}

Another possible shaping mechanism is that the $B$ field lines together with the dust grains are 
dragged along with the outflows (Figure 7(b)). In this case $-$ in order to explain the 
polarization patterns perpendicular to the outflows (Section 4.2) $-$ the dust grains are required to remain 
anchored to the $B$ field in an environment where the bulk motions are substantial and dramatic, so 
that a mechanical dust grain alignment might be possible. The velocity of the outflow is up to 100 
km s$^{-1}$, which is highly supersonic. However, if the explosive outflows drag along the $B$ field lines, 
dust grains, and other molecules all together, the dust grains in the local rest frame might not suffer 
from a higher collision rate and therefore can still remain magnetically aligned. Furthermore, the 
kinematics within the clumps, inferred from the narrow linewidths of NH$_{3}$ lines, indicates that the 
clumps do not have large internal velocity dispersions. This is consistent with the suggestion that 
the outflow motions affect mainly the bulk motions and also large$-$scale $B$ field lines. It remains to 
be determined whether the outflow energy density (pressure) is enough to drag and shape the $B$ 
field lines. 

The $B$ field tension, $f_{\rm B}$, can be approximately estimated by using the equation: 
$f_{\rm B}$ = $\frac{B^{2}}{4\pi R}$. The $B$ field strength, $B$, near IRc2 is in the range 1 mG$-$16 mG, based 
on the OH maser Zeeman splitting (Johnston, Migenes, \& Norris 1989; Cohen et al. 2006) and is 0.4 mG along 
the line of sight based on the Zeeman splitting of the CN line (Crutcher et al. 1999). Note that $B$ estimated to 
confine the NH$_{3}$ clumps (Section 4.1.2) is $\ge$ 3 mG if $n_{\rm H_2}$ of the outflows is $\sim$ 10$^{-6}$ cm
$^{-3}$, which is consistent with the OH maser measurements. The curvature of the $B$ field, $1/R$, is difficult to 
estimate because of the azimuthal symmetry and also the existence of a preferred symmetry plane. Here, we 
adopt a curvature of 
1/10$\arcsec$ based on the change in P.A.s along the dust ridge. The estimated $f_{\rm B}$ is then in the range of 1 
$\times$ 10$^{-24}$ to 3 $\times$ 10$^{-22}$ (dyn cm$^{-3}$). The momentum transfer rate per unit volume, 
$f_{\rm outflow}$, can be estimated from 8 $\times$ 10$^{6}$ L$_{\sun}$ c$^{-1}$ (Kwan \& Scoville 1976) divided 
by the volume of interest, where $c$ is the speed of light. Since the large$-$scale outflows have a common center near the BN-I center, it is reasonable to assume that the 
energy contained in the outflows originates in the BN-I center. Assuming that the energy 
released in the outflows is uniformly distributed over a sphere of radius of 10$\arcsec$, $f_{\rm outflow}$ is 4 $
\times$ 10$^{-15}$ (dyn cm$^{-3}$). Even assuming a volume with a radius 10 times larger, which would cover 
the entire outflow area, would still give $f_{\rm outflow}$ $\approx$ 4 $\times$ 10$^{-18}$ (dyn cm$^{-3}$). 
Although $f_{\rm outflow}$ depends on the assumed age of the outflow, where 3000 yr is used in 
Kwan \& Scoville (1976), the fact that $f_{\rm outflow}$ is at least 4 orders of magnitude larger than 
$f_{\rm B}$ suggests that the outflows are energetic enough to distort the $B$ field lines of curvature of 1/10$\arcsec$. However, it is extremely difficult to shape the field lines because in the shaping process, both $1/R$ and $B$ increase. This scenario is more likely if such a distortion happens as bulk motion, i.e., $R$ will not 
decrease infinitely but has a minimum of 0.5 AU, which is 10$^{-4}$ of 10$\arcsec$, if $B$ remains as 16 mG in 
the process.

In order to distinguish between the driving mechanisms for the shaped field lines, observations with a more complete $uv$ coverage, higher angular resolution, and higher sensitivities are needed. The link of the field geometry near the BN-I center and the locations where no polarization was detected will be helpful. ALMA will be able to improve this picture. For example, if the field lines are shaped by explosive outflows, an azimuthal symmetry of $B$ field lines 
is expected to point all the way back toward the BN-I center, where there is de-polarization in current 
observational results. If the field lines are shaped by rotation or accretion, both the toroidal field lines, which are connected to the magnetized cores, and the poloidal fields from large$-$scale fields are expected. The polarization emission from different molecular lines via the Goldreich$-$Kylafis mechanism (Goldreich \& Kylafis 1981) might also be helpful to trace the $B$ field geometry in different physical conditions, e.g., lower density regions. Besides observations, analysis tools to quantify the mirror and radial symmetry will be needed.

\subsection{Comparison with other massive star formation sites}
The $B$ field geometries in the collapsing cores of the massive star formation regions W51 e2/e8 
(Tang et al. 2009) and G31.41 (Girart et al. 2009) are found to be highly pinched, similar to what 
has been found in the low mass cases. We note that both W51 e2/e8 and G31.41 are at larger 
distances of 7 kpc and 7.8 kpc, respectively. The signature of rotation has been detected with the 
recombination line in W51 e2/e8 and with molecular lines in G31.41. The radius of the rotation is 
$\sim$1$\arcsec$. If these sources were at the distance of 480 pc as Orion BN/KL, the scale of the rotation 
signature would be up to 16$\arcsec$, comparable to the scale of the distorted $B$ field lines presented here.
 
A uniform $B$ field on the scale of 0.5 pc is found in W51 e2/e8 with two collapsing cores 
inside a larger scale envelope. Tang et al. (2009) have proposed that these are most likely due to 
magnetic fragmentation. In OMC-1, we find that the structures seen in the W51 e2/e8 region are 
possibly not a unique case. At a 0.5 pc scale, the $B$ field is also uniform, and there are dense cores 
(Orion BN/KL and Orion south) detected. The comparison with star$-$forming cores within stable 
envelopes is important to test different fragmentation scenarios. When observed with higher angular resolution, the field lines within dense cores in both regions appear to be shaped by either gravity or outflows. This might be a hint of the changing role of the $B$ field as a function of scale. To test the dependence of scale, we have started a statistical analysis with polarization measurements combining data from different scales. See Koch, Tang \& Ho (2010) for more details. 

Observational and theoretical predictions indicate that the ambipolar diffusion scale is on the 
order of 1 mpc. In this paper, we are able to measure the $B$ field morphology close to this turbulence 
scale. Based on the smooth polarization morphology, we conclude that the $B$ field is also dominant 
over turbulence close to the scale of 2 mpc. This is consistent with previous observations in M17 
by Li \& Houde (2008). As compared to gravity and the large$-$scale kinematics, the B field is clearly 
overwhelmed by these two forces at the 2 milli-pc scale. Even though the $B$ field lines are highly shaped by 
the large scale kinematics, the overall B field morphology is still organized and not affected much 
by the small-scale turbulence, if they are present.

\section{Summary and Conclusion}
The 870 $\mu$m dust continuum and its linearly polarized emission
are detected and resolved in the Orion BN/KL region. With an
angular resolution of 1$\arcsec$, it is the first time that the polarized emission from the
dust continuum at submillimeter wavelengths in the Orion region is resolved at 2 mpc (480 AU), which is about the ambipolar diffusion scale
in M17. As compared to the dust
polarization previously detected, the revealed polarization in the
south of IRc2 changes smoothly by 90$\degr$ instead of abruptly as previously reported at lower angular resolution. The continuum
emission appears to be $V$-shaped with a cavity near the center of the explosive outflows (Section 3.1), called  the BN-I center in this paper. The brightest dust clump is associated with source I and SMA 1.
  
The detected polarization appears to be curl-like near the BN-I center (Section 3.2). Such a polarization pattern indicates that the dust grains are most likely magnetically aligned instead of mechanically aligned with the outflows because the latter mechanism would produce a radial-like polarization pattern (Section 4.2). The magnetic ($B$) field projected in the plane of sky is therefore derived by rotating the P.A.s of the 
polarization by 90$\degr$.

We have checked the symmetry properties of the inferred $B$ field lines. There is an azimuthal symmetry of the
$B$ field lines centered about 2$\farcs$5 west of the BN-I center (Section 4.3), indicating that the $B$ field pattern might be correlated with the explosive outflows. Besides the azimuthal symmetry, there are polarization patches which clearly deviate from a radial symmetry. Therefore, we have further checked if there is additionally a mirror symmetry in the detected $B$ field lines. Indeed, there is a preferred plane of symmetry 
at a P.A. of 36$\degr$, nearly perpendicular to the large$-$scale $B$ field direction at a P.A. of 120$\degr$. This is expected in the formation of a dense core within a magnetized molecular cloud, where most material can flow along the field lines and thus dense structures form perpendicular to the field direction.

One possible interpretation of the detected $B$ field lines is that there are both poloidal and toroidal fields near the BN/KL region, possibly associated with a remnant pseudo-disk (Section 4.3.1). In this scenario, the detected $B$ field inferred from the dust continuum 
is toroidal in the south of source I and east of IRc5, and is mostly poloidal in the rest of the detected field lines. 
The poloidal $B$ field is consistent with the field direction observed at 0.5 pc scale in the OMC-1 dust ridge. Since the estimated period of such a remnant pseudo-disk is 1$-$2 orders of magnitude shorter than the typical star formation time scale, mechanisms to prevent the field lines from 
winding up are needed. Another interpretation of the detected $B$ field is that the field lines are dragged along with the 0.14 
pc scale, explosive CO outflows (Section 4.3.2). The momentum transfer rate of the outflow is at least 4 orders of magnitude larger than the $B$ field tension with a curvature of 10$\arcsec$, indicating that the outflows are energetic enough to distort the field lines. However, it will require much more energy to distort the field lines if the curvature of the field lines decreases and if the field strength increases due to flux conservation. This scenario is more likely if such distortion happens as bulk motion.

In order to distinguish between these two scenarios, it will be very helpful to obtain the field geometry of the 
depolarization region near the BN-I center and of the lower density regions. Observations with higher angular 
resolution and sensitivity are required. ALMA will be able to provide much more sensitive and higher angular resolution images. Besides dust polarization, measurements of molecular line polarization also might be helpful to probe the $B$ field geometry in different physical conditions.
The analysis tools to quantify the mirror symmetry and compare it to the radial symmetry will also be 
helpful to further solve the mystery.    
We note that in the main dust ridge, it has been suggested that source I and SMA are newly formed stars, indicating that further fragmentation is ongoing within the remnant disk.

The authors acknowledge the anonymous referee for the helpful comments, which have improved the presentation of the paper. Y.-W. T. is grateful to Josep M. Girart and Mike Cai for the interesting discussions. Y.-W. T. and P. T. P. H. are supported by the NSC grant NSC97-2112-M-001-007-MY3 and NSC98-2119-M-001-024-MY4.

\begin{deluxetable}{ccccc}
\tablecaption{Observational parameters} \tablewidth{0pt}
\tablehead{ \colhead{Date} & \colhead{Configuration} &
\colhead{Gain Calib.}& \colhead{Flux Calib.} & \colhead{BP/Pol.
Calib.}}

\startdata

2008 Jan 6 & Compact & 0528+134 & Uranus & 3c273 \\
2009 Jan 25 & Subcompact & 0423$-$013/0510+180 & Titan & 3c273 \\
2009 Feb 15 & Extended & 0423$-$013 & Titan & 3c273 \\

\enddata
\tablecomments{BP/Pol. refers to bandpass/polarization. Calib.
refers to the calibrator. In all of the observations, the local
oscillator frequency was tuned at 341.482 GHz. In both compact
and subcompact tracks, seven of eight antennas were available. In the
extended track, there were eight antennas available.}
\end{deluxetable}

\newpage
\begin{figure}
\includegraphics[angle=0,scale=0.7]{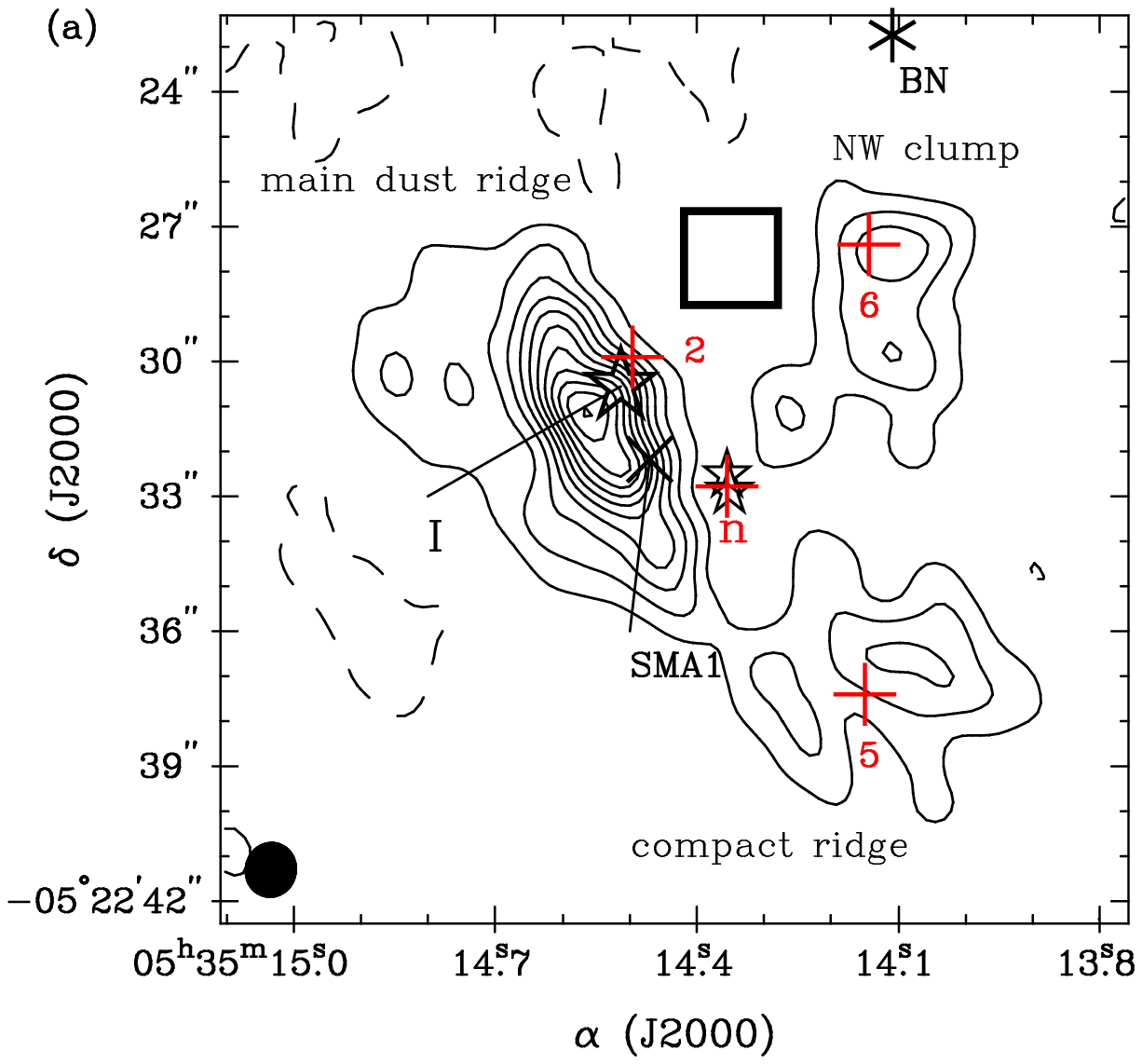}
\includegraphics[scale=0.7]{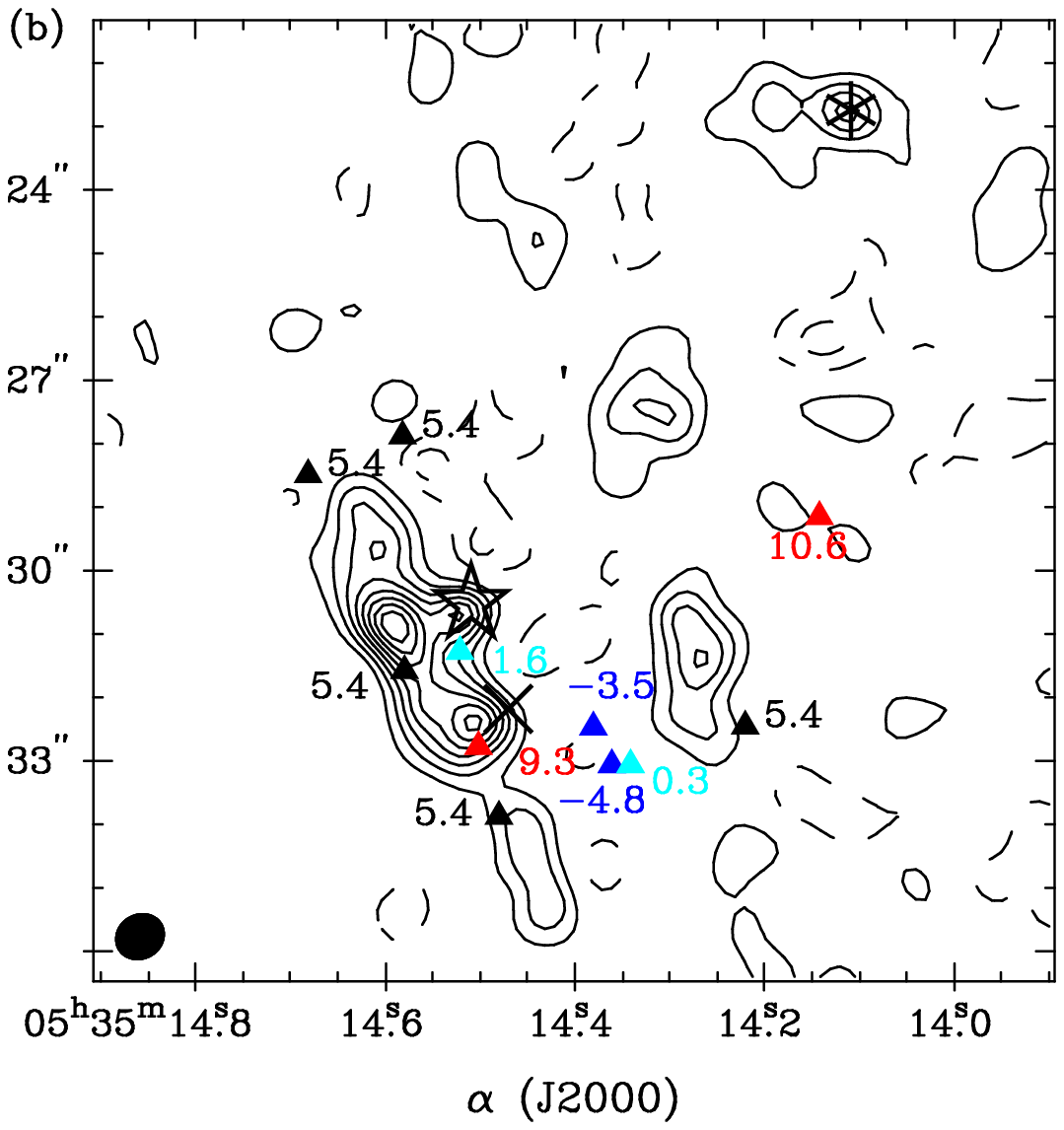}
\caption{SMA 870 $\mu$m dust continuum maps. (a) Map obtained from
combined three tracks with natural weighting. The synthesized beam ($\theta_{\rm syn}$)
is 1$\farcs$2$\times$1$\farcs$1 with a P.A. of $-$13$\degr$, shown as solid black ellipse at the lower-left corner.
Solid contours start from and step in 3$\sigma_{\rm I}$, where $\sigma_{\rm I}$ is the noise level of Stoke I  image and is 90 mJy beam$^{-1}$ here. Dashed contours are $-$3$\sigma_{\rm I}$, which is due to the missing of short spacing visibilities. The asterisk, the stars, and the cross mark the position of BN, the radio sources identified by Menten \& Reid (1995), and the
submillimeter continuum source SMA1 identified by Beuther et al.
(2004), respectively. Red pluses are the mid-infrared sources,
where the adjacent numbers refer to the IRc source names in
Shuping, Morris \& Bally (2004).  The black square marks the BN-I center. 
(b) Map obtained from extended track with natural weighting. The $\theta_{\rm syn}$ is 
0$\farcs$8$\times$0$\farcs$7 with a P.A. of $-$59$\degr$. The triangles mark the positions of clumps identified
in NH$_{3}$ by Migenes et al. (1989) with $v_{\rm lsr}$ labeled
in units of km s$^{-1}$. The dashed and solid contours are plotted in the same steps as in panel (a), except that 
$\sigma_{\rm I}$ is 40 mJy beam$^{-1}$ here.}
\label{cont_ip}
\end{figure}

\begin{figure}
\includegraphics[scale=0.72]{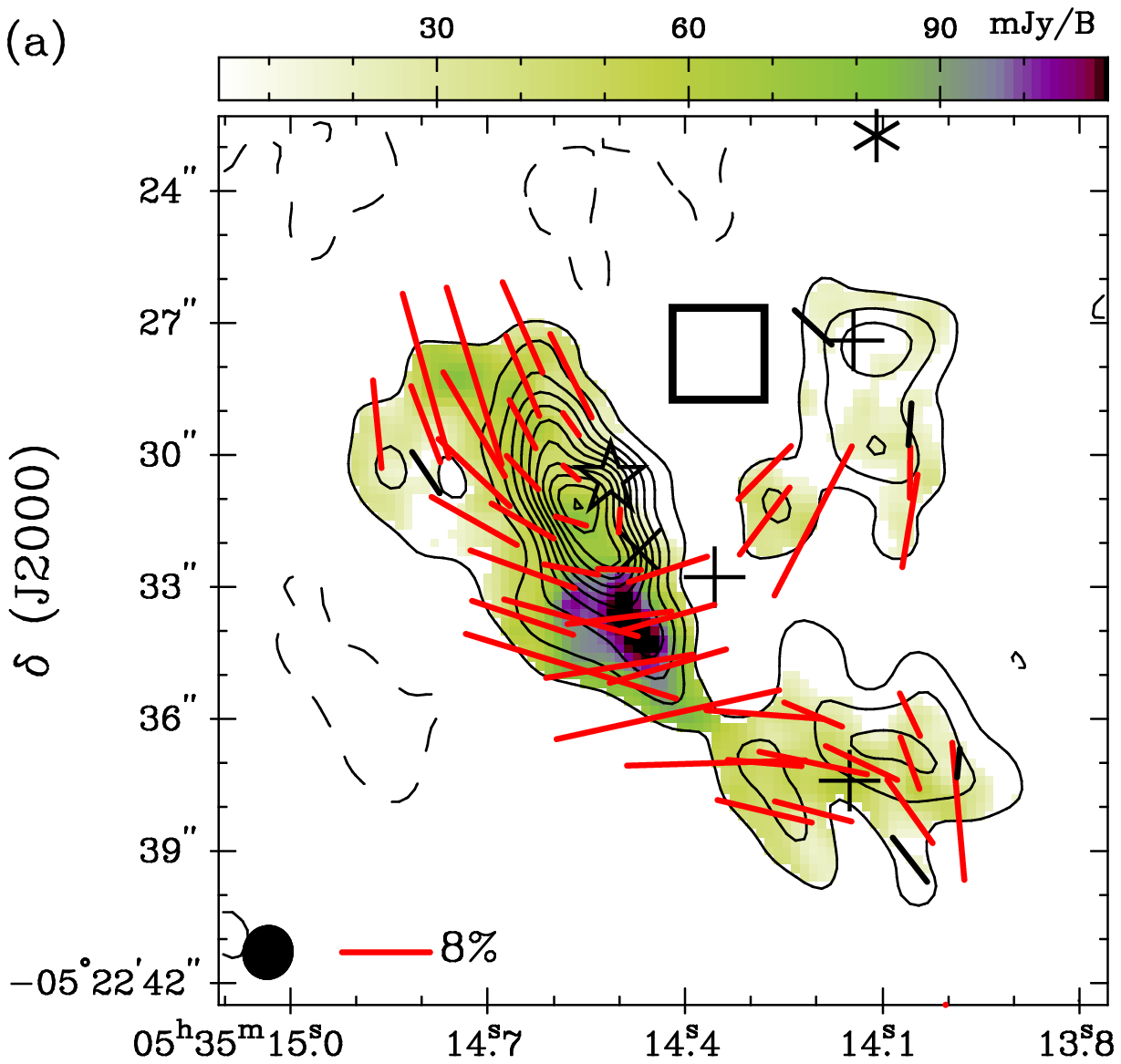} 
\includegraphics[scale=0.72]{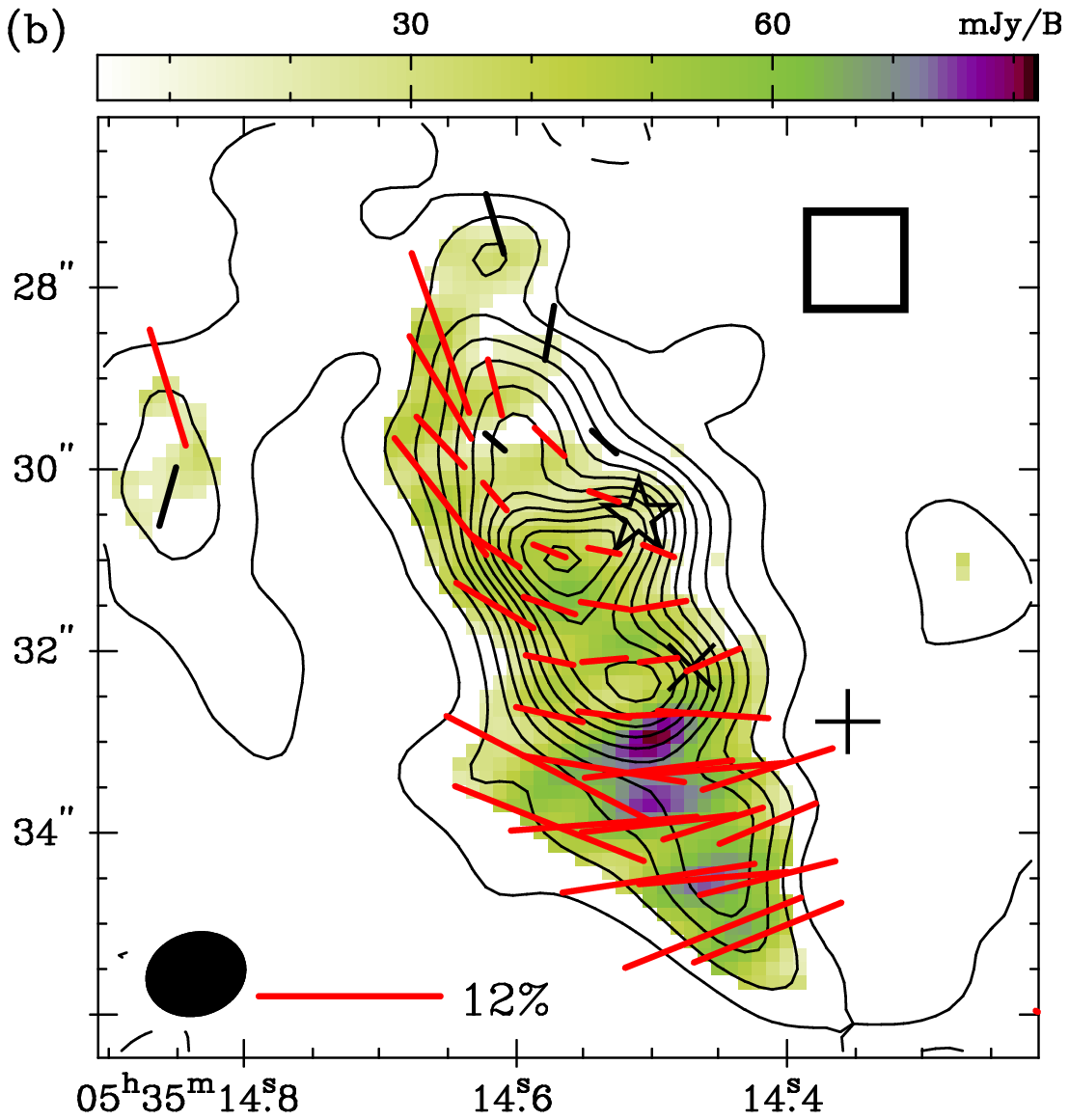} 
\includegraphics[scale=0.72]{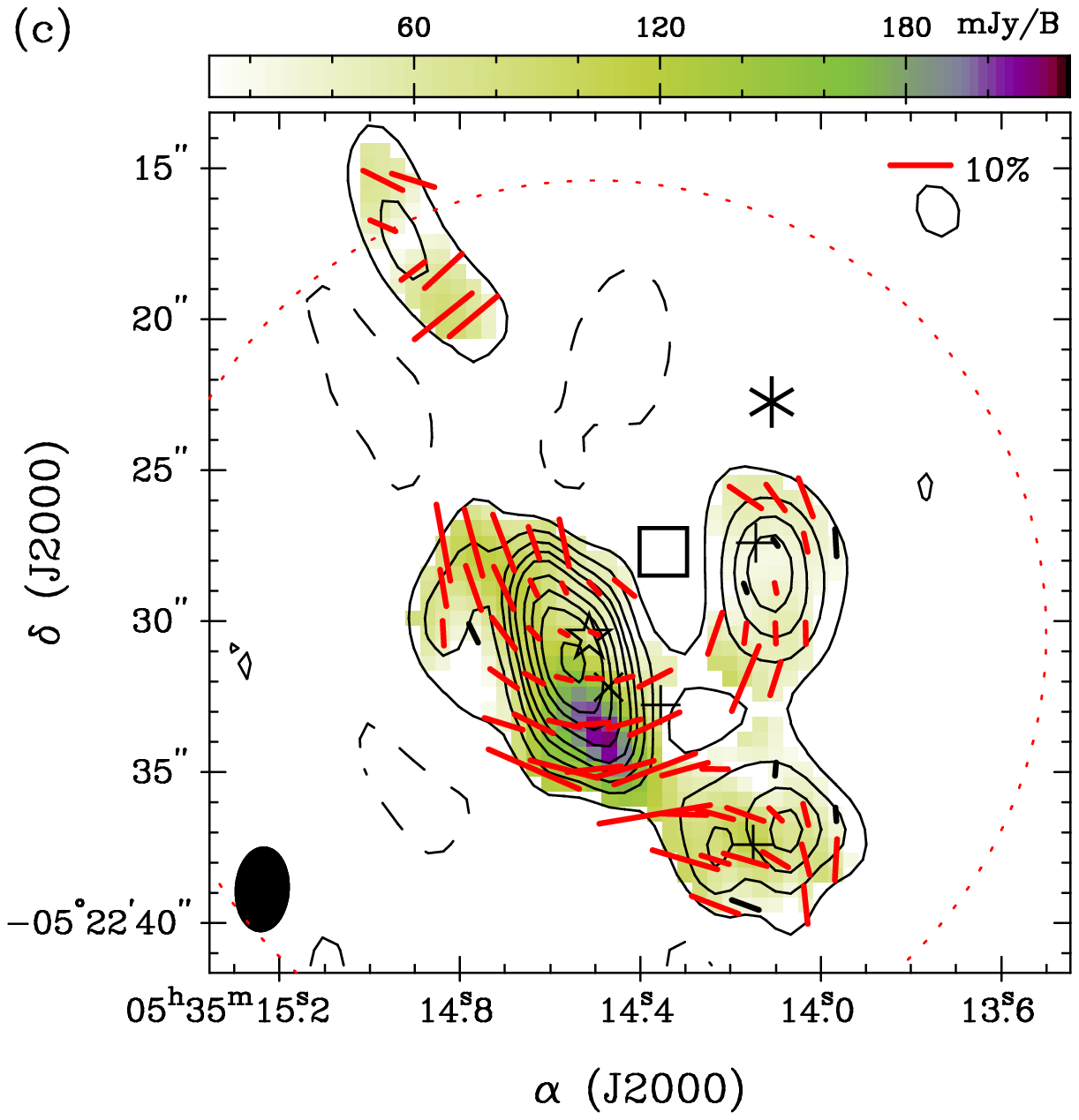} 
\includegraphics[scale=0.72]{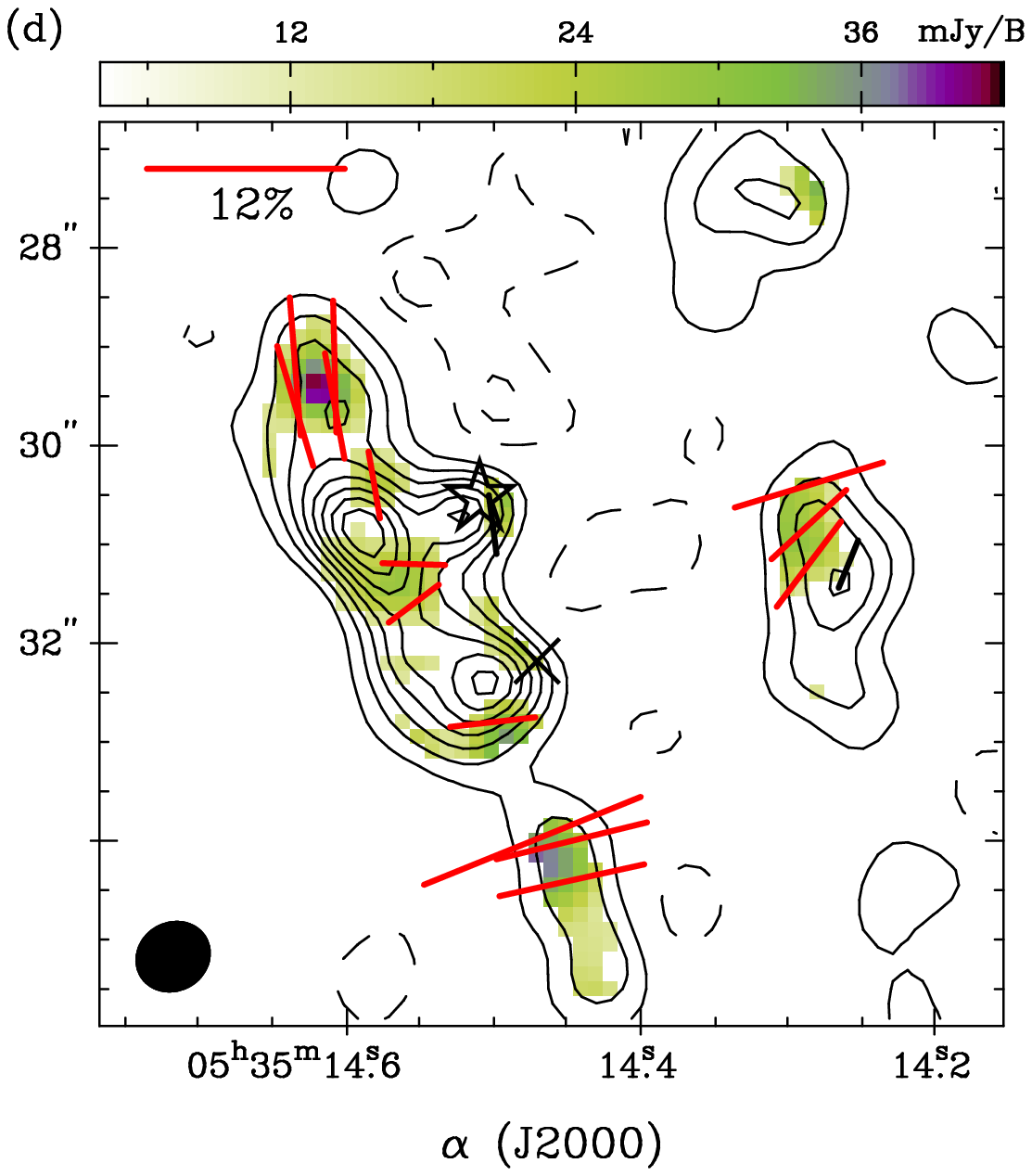}
\caption{\small{Polarized intensity (color scale), polarization
(segments) and 870 $\mu$m dust continuum emission (contours) maps obtained with different data sets and weightings. The red and black segments are above 3 $\sigma_{\rm p}$ and between 2 and 3
$\sigma_{\rm p}$, respectively, where $\sigma_{\rm p}$ is the noise level of the polarized intensity. The symbols are the same as in
Figure 1. (a) Three tracks combined with natural weighting. Contour
levels are the same as in Figure 1(a). Polarization segments are gridded with 1$\farcs$1. (b) Map obtained with the combined three tracks with uniform weighting. $\theta_{\rm syn}$ is 1$\farcs$1$\times$0$\farcs$9 with P.A. of $-$73$\degr$. Dashed and solid contours are plotted as $-$6, $-$3, 3, 6, 9, ..., 30, 33, 36 $\times$ 50 mJy beam$^{-1}$. The vectors are gridded with 0$\farcs$6 in order to show the variation of
polarization across each independent data point. (c) Combined compact and subcompact
array data. With robust weighting of 0.5, the synthesized beam is
2$\farcs$8$\times$1$\farcs$8 with P.A. of -2$\degr$. The contours are plotted as $-$3, 3, 6, 9, 12, 15, 20, 25, 30, 35,
40 $\times$ 0.16 Jy beam$^{-1}$. Polarization segments are gridded with 1$\arcsec$ in R.A. and 1$\farcs$5 in decl. The dotted circle marks the primary
beam of the SMA at this wavelength. (d) Map obtained with the
extended array track with natural weighting. Polarization segments are gridded with 0$\farcs$4. Contour levels and $\theta_{\rm syn}$ are
the same as in Figure 1(b).}} \label{ip_pol_3tra}
\end{figure}

\newpage
\begin{figure}
\includegraphics[scale=0.5]{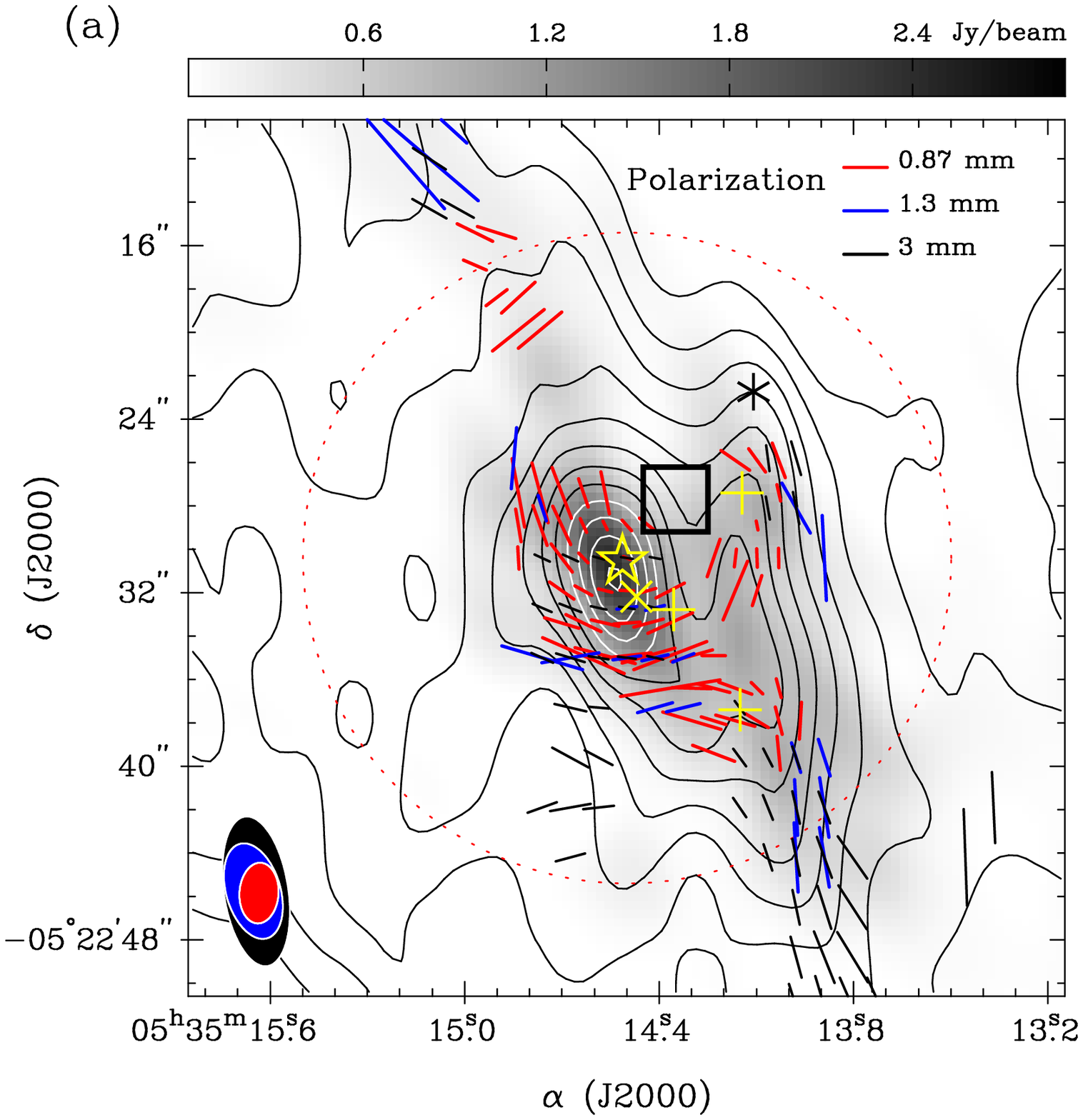}
\includegraphics[scale=0.7]{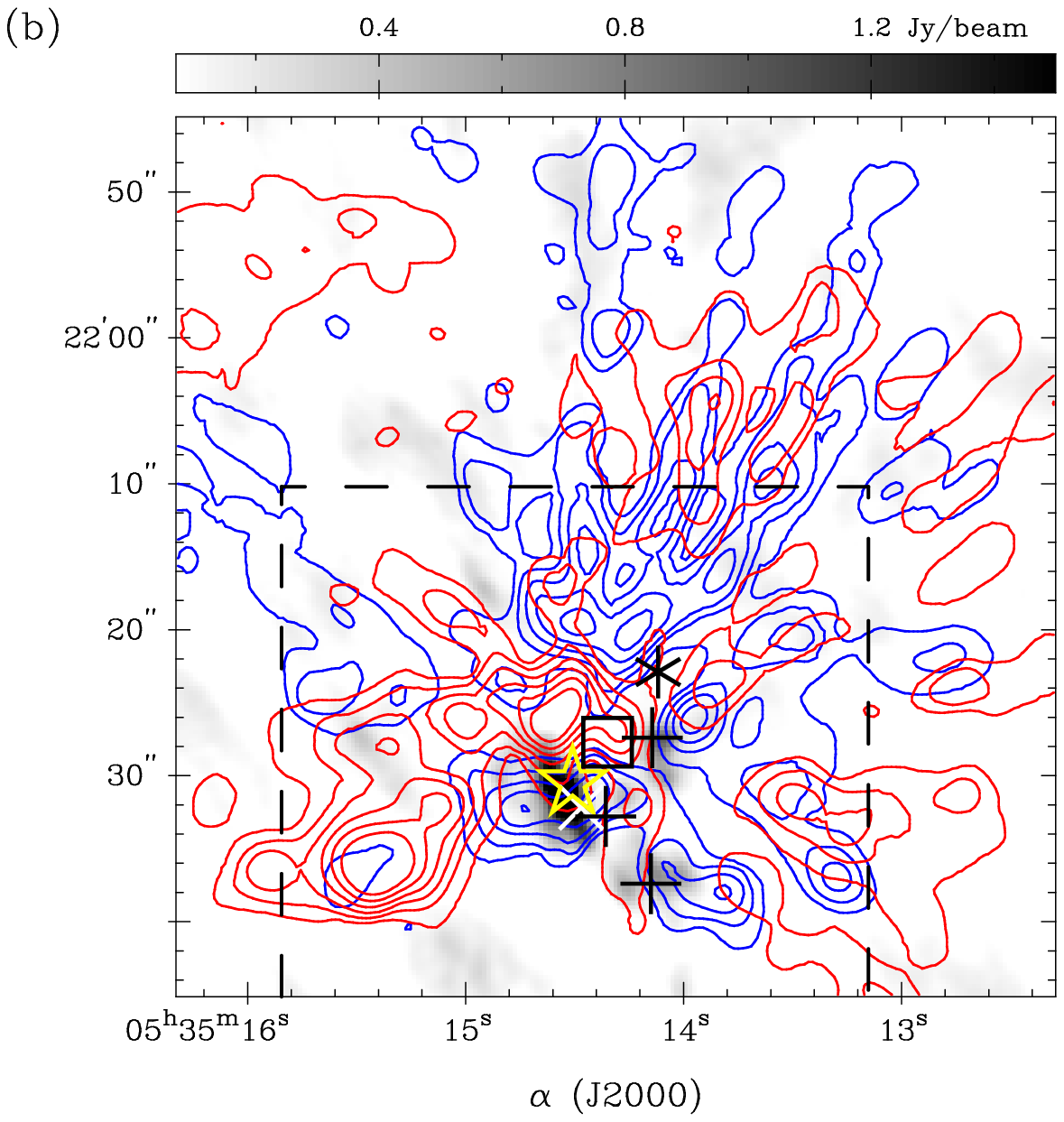}
\caption{(a) Polarization maps at 3 mm (black segments) and 1 mm
(blue segments) obtained with BIMA (Rao et al. 1998) and at 870 $\mu$m
(red segments; the same as in Figure 2(c)) obtained with the SMA. The black contours are the 3 mm
continuum emission strength at 3, 6, 9, ..., 33, 36, 39 $\times$
0.01 Jy beam$^{-1}$. The 1 mm continuum emission is shown in grayscale. The sizes of the synthesized beams are plotted in the
lower-left corner in the corresponding color as indicated in the
upper-right corner. The length of the indicated segments in these
three wavelengths represents the polarization percentage of 8\%.
The large circle in red dots marks the field of view of the
SMA. All the other symbols are the same as in Figure
1(a). (b) 870 $\mu$m dust continuum (grayscale) of the combined
tracks with natural weighting and CO outflows in blue and red
contours by Zapata et al. (2009). The contours
start from 8 and step in 24 Jy beam$^{-1}$ km $^{-1}$. The black dash
square marks the presented region in panel (a).}
\label{B_bima_sma}
\end{figure}

\newpage
\begin{figure}
\includegraphics[scale=0.7]{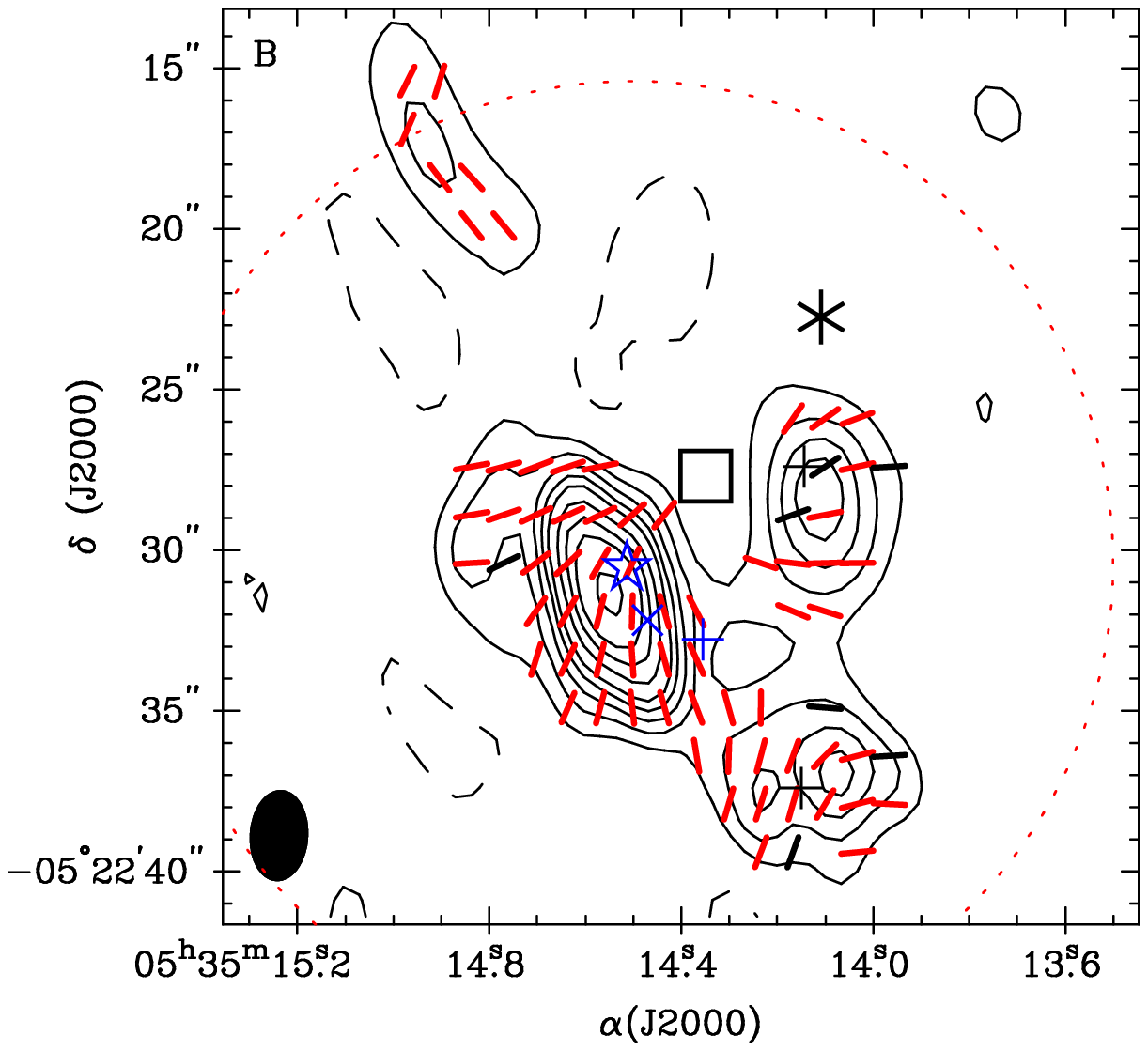}
\includegraphics[scale=0.7]{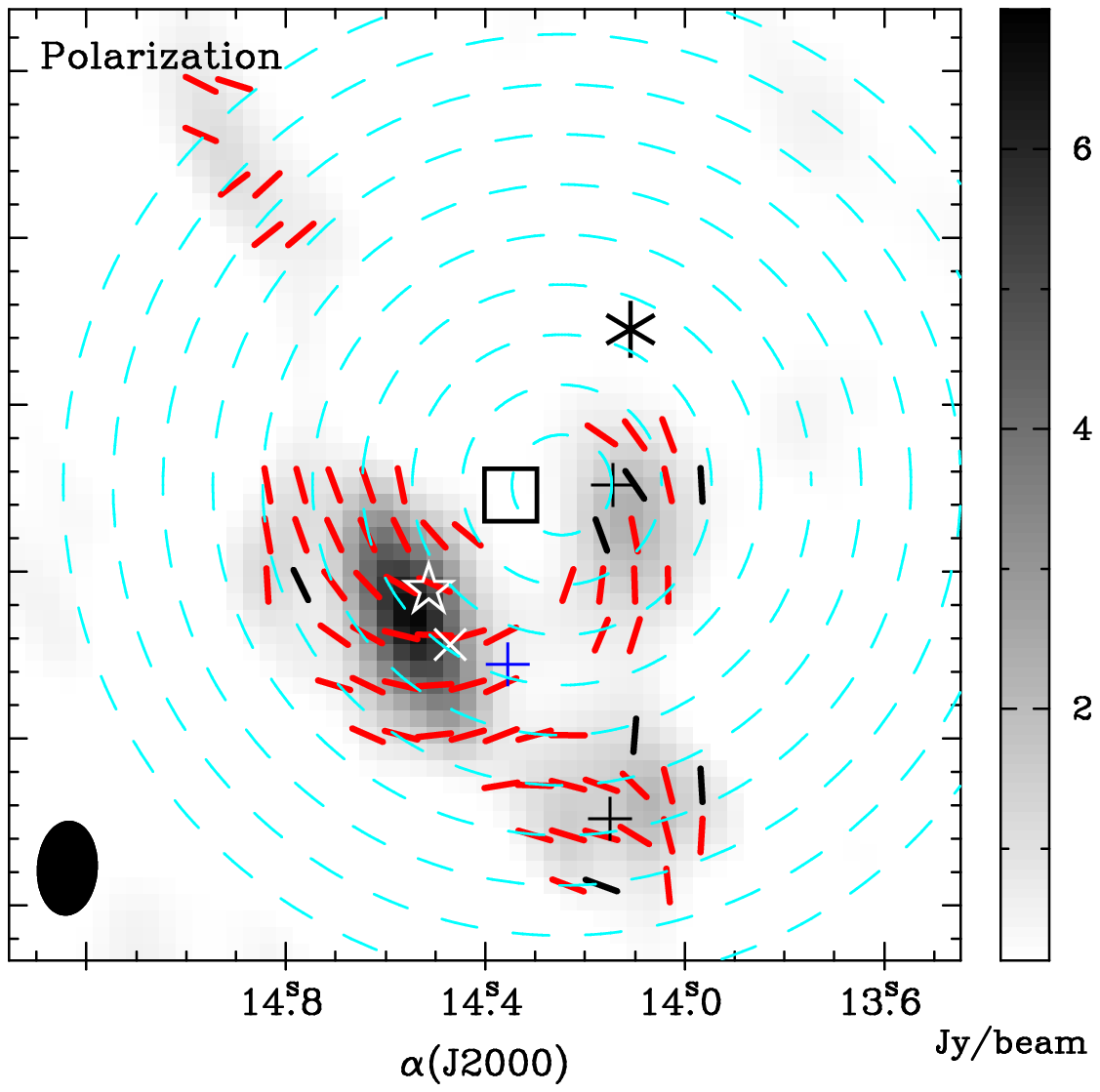}
\caption{Left panel: $B$ field map (red and black segments) obtained with the combined compact and subcompact data. The $B$ field vectors are derived by rotating the polarization segments (Figure 3) by 90$\degr$ with identical length. Right panel: polarization map (red and black segments) and 870 $\mu$m dust continuum (grayscale) from the combined compact and subcompact data. The concentric circles in cyan dashes are centered on the best azimuth symmetry center, which is 2$\farcs$5 west to the BN-I center. The other symbols are the same as in Figure 2(c).} \label{cont_ip}
\end{figure}

\newpage
\begin{figure}
\includegraphics[scale=0.8]{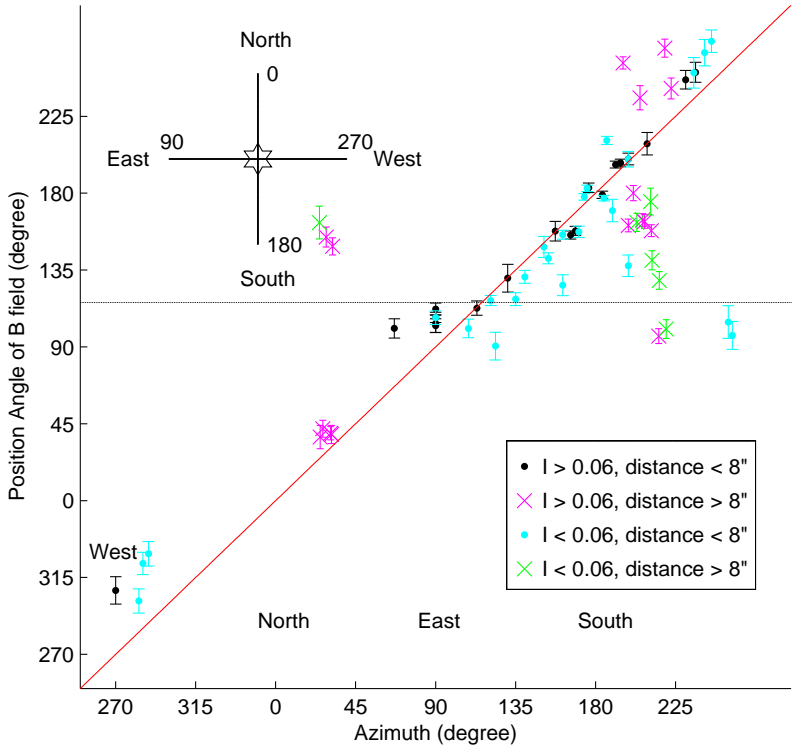}
\caption{Azimuth positions of detected $B$-field directions in the left panel of Figure 4. 
The reference origin is 2$\farcs$5 west to the BN-I center. The red line marks the expected P.A.s for a purely radial field. The black solid line marks the mean P.A. of the $B$ field at 0.5 pc scale (Vall\'{e}e 
\& Fiege 2007). Data points are color coded with the relative distance to the origin in units of 
arcsecond for two different intensity levels (labeled as \textsf{I} in units of Jy beam$^{-1}$ in the legend). See 
Section 4.3 for details.}
\label{cont_ip}
\end{figure}

\newpage
\begin{figure}
\includegraphics[scale=0.7]{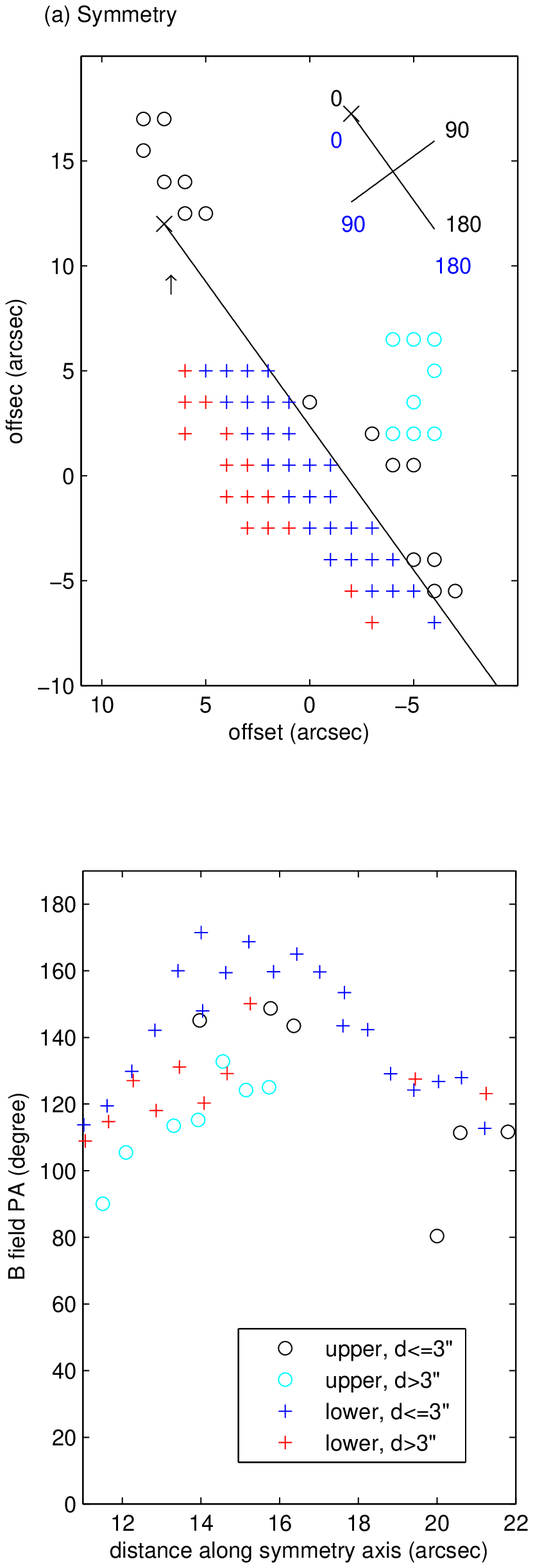}
\includegraphics[scale=0.7]{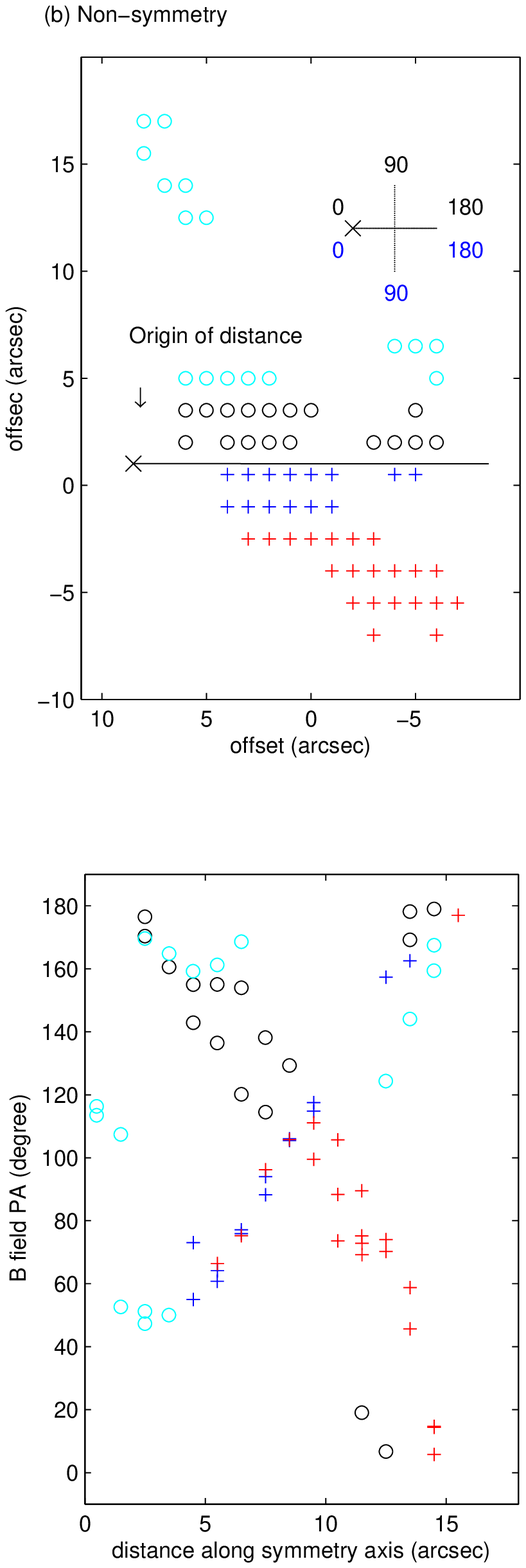}
\caption{\small{Plots along the best-symmetry plane at a P.A. of 36$\degr$ (left panels) and along 
a horizontal plane (right panels). The detected polarizations are separated into two groups: data 
above the plane (marked as circles) and data beneath the plane (marked as pluses). The symmetry 
plane is marked as a black line. The origins of distance along the symmetry axes are marked as black crosses in the upper panels. In the upper panels, the offsets are relative to the phase center. Data points are further color coded with the relative distance to the test planes (labeled as \textsf{d}) with a separation of 3$\arcsec$.} The $B$ field P.A.s are redefined from 0$\degr$ to 180$\degr$, increasing clockwise 
for the upper plane and counterclockwise for the lower plane, as shown in the legend. The distribution of the corresponding redefined P.A.s are shown in the lower panels.} \label{cont_ip}
\end{figure}

\newpage
\begin{figure}
\includegraphics[scale=0.5]{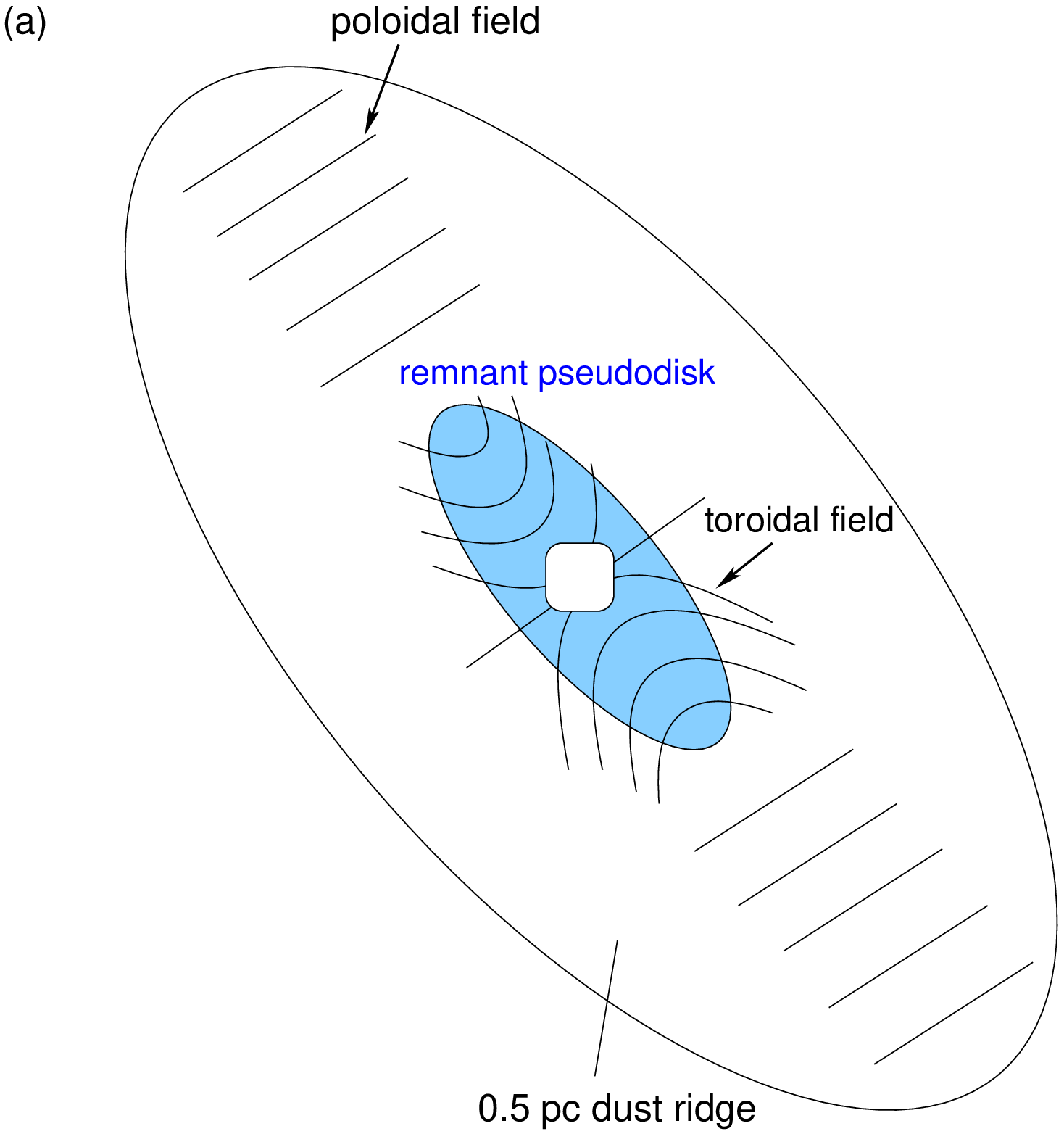}
\includegraphics[scale=0.5]{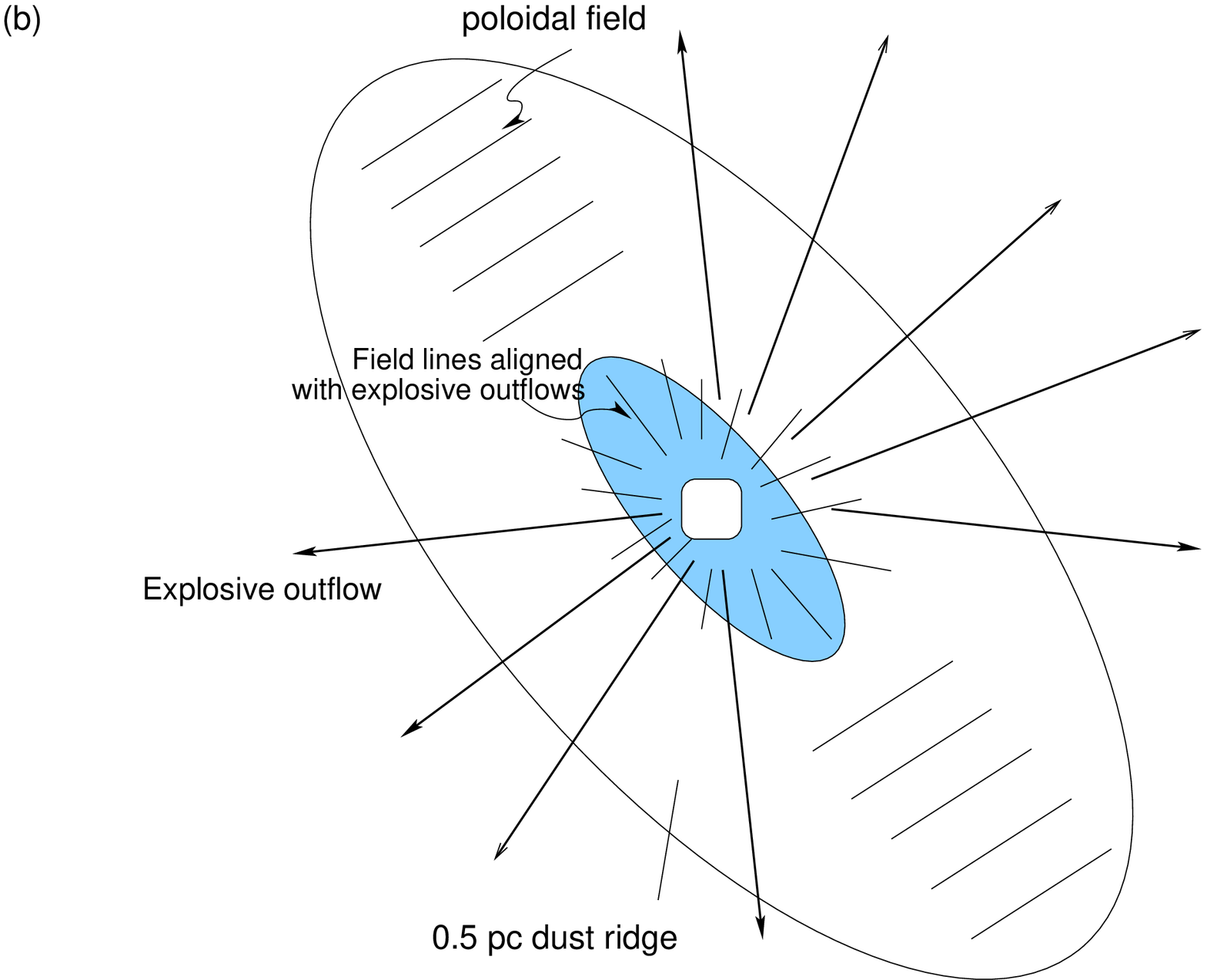}
\caption{Schematics of two possible interpretations. (a) $B$ field
lines in the pseudo-disk are nearly toroidal due to the
differential rotation of the magnetized disk. (b) $B$ field lines
are dragged along with the explosive molecular outflow. See Section 4.3
for more discussions.} \label{schematic}
\end{figure}

\end{document}